\documentclass[a4paper,10pt,aps,prl,longbibliography,twocolumn,amsmath,amssymb,nofootinbib,superscriptaddress]{revtex4-2}
\usepackage{graphicx}

\usepackage[colorlinks=true,linkcolor=blue,citecolor=blue]{hyperref}
\usepackage{amsmath,bm,epsfig,color}
\usepackage{floatrow}
\usepackage{epsfig}
\usepackage{bm}

\usepackage[latin1]{inputenc}
\renewcommand{\=}{\!=\!}

\usepackage{xcolor,hyperref}
\hypersetup{
   colorlinks,
   linkcolor={blue!50!black},
   citecolor={blue!50!black},
   urlcolor={blue!80!black}
}

\begin{document}
\title{Bridging necking and shear-banding mediated tensile failure in glasses}

\author{David Richard}
\thanks{Contributed equally}
\affiliation{Univ. Grenoble Alpes, CNRS, LIPhy, 38000 Grenoble, France}
\author{Ethen Lund}
\thanks{Contributed equally}
\affiliation{Department of Mechanical Engineering and Materials Science, Yale University, New Haven, CT, USA}
\author{Jan Schroers}
\email{jan.schroers@yale.edu}
\affiliation{Department of Mechanical Engineering and Materials Science, Yale University, New Haven, CT, USA}
\author{Eran Bouchbinder}
\email{eran.bouchbinder@weizmann.ac.il}
\affiliation{Chemical and Biological Physics Department, Weizmann Institute of Science, Rehovot 7610001, Israel}

\begin{abstract}
The transition between necking-mediated tensile failure of glasses, at elevated temperatures and/or low strain-rates, and shear-banding-mediated tensile failure, at low temperatures and/or high strain-rates, is investigated using tensile experiments on metallic glasses and atomistic simulations. We experimentally and simulationally show that this transition occurs through a sequence of macroscopic failure patterns, parametrized by the ultimate tensile strength. Quantitatively analyzing the spatiotemporal dynamics preceding failure, using large scale atomistic simulations corroborated by experimental fractography, reveals how the collective evolution and mutual interaction of shear-driven plasticity and dilation-driven void formation (cavitation) control the various macroscopic failure modes. In particular, we find that at global failure, the size of the largest cavity in the loading direction exhibits a nonmonotonic dependence on the temperature at a fixed strain rate, which is rationalized in terms of the interplay between shear- and dilation-driven plasticity. We also find that the size of the largest cavity scales with the cross-sectional area of the undeformed sample. These results shed light on tensile failure of glasses, and highlight the need to develop elasto-plastic constitutive models of glasses incorporating both shear- and dilation-driven irreversible processes.
\end{abstract}

\maketitle


Glasses subjected to sufficiently large tensile stresses, like any other materials, inevitably fail. At elevated temperatures and/or low strain-rates, deformation is essentially homogeneous and failure is known to be mediated by necking~\cite{spaepen1977microscopic,lu2003deformation}, where spatially-extended plastic deformation geometrically localizes at a radially-symmetric, shrinking cross-sectional area~\cite{wang2005tensile,vormelker2008effects}. At low temperatures and/or high strain-rates, deformation is highly localized and failure is mediated by shear-banding~\cite{spaepen1977microscopic}, where plastic deformation strongly localizes at an oblique (symmetry-breaking) plane, before transforming into a catastrophic crack~\cite{wang2005tensile,vormelker2008effects}. While progress has been made in understanding these two end-members of tensile failure modes of glasses, our understanding of the transition between them as a function of the temperature and strain-rate significantly lags behind. In this study, using tensile experiments performed on well-controlled metallic glasses and large-scale molecular dynamics simulations of model glasses, we investigate the transition between necking- and shear-banding-mediated tensile failure in glasses. We focus both on the macroscopic tensile failure patterns and on the spatiotemporal elasto-plastic dynamics that accompany them, paying special attention to the interplay between shear-driven plasticity and dilation-driven void formation (cavitation)~\cite{schroers2004ductile,bouchbinder2007stability,bouchaud2008fracture,jiang2008energy,bouchbinder2008dynamic}.

\begin{figure*}[t!]
\includegraphics[width = \textwidth]{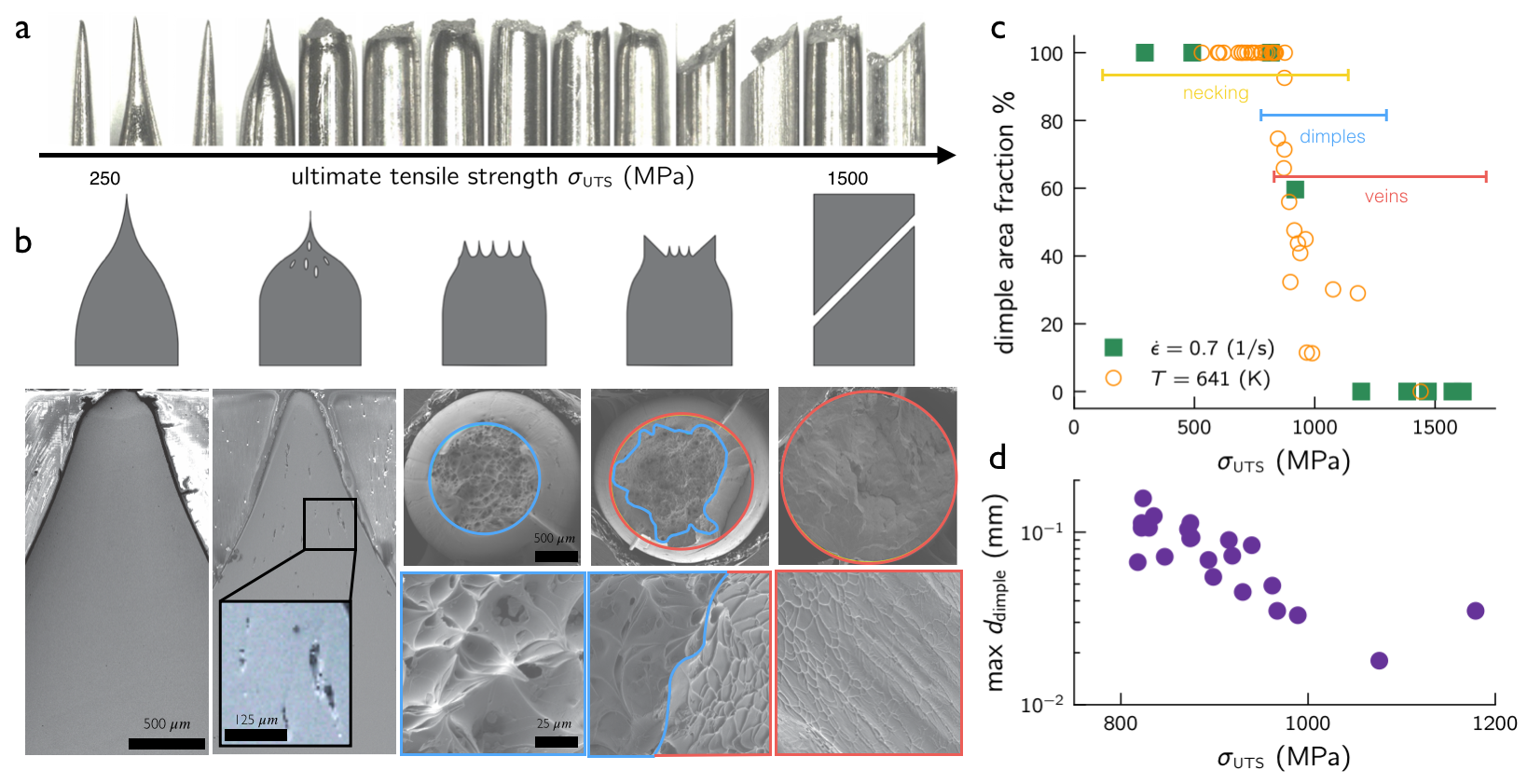}
\caption{(a) Side view of the failed samples sorted by increasing ultimate tensile stress $\sigma_{\mbox{\tiny UTS}}$, which is varied either by varying $T$ or $\dot\epsilon$. $\sigma_{\mbox{\tiny UTS}}$ varies by roughly a factor of 6.5, approximately from 250 MPa to 1600 MPa as indicated, but note that the presented samples are not placed at equal $\sigma_{\mbox{\tiny UTS}}$ intervals. (b) Sketches of the macroscopic failure modes (top row), ranging from necking-mediated failure on the left (low $\sigma_{\mbox{\tiny UTS}}$, corresponding to high $T$ or low $\dot\epsilon$) to shear-banding-mediated failure of the right (high $\sigma_{\mbox{\tiny UTS}}$, corresponding to low $T$ or high $\dot\epsilon$), see text for details. (bottom row) Vertical sample polishing (two leftmost parts) and fractographic images (the rest), corresponding to the sketches above (scale bars are added), see extensive discussion in the text. The stated scale bars apply to each set of adjacent micrographs, respectively. (c) Dimple area fraction vs.~$\sigma_{\mbox{\tiny UTS}}$, for a fixed $\dot\epsilon$ and varying $T$ (green squares) and a fixed $T$ and varying $\dot\epsilon$ (orange circles), see legend. The range of observed necking, dimples, and veins is indicated by horizontal colored bars. (d) The maximum dimple size vs.~$\sigma_{\mbox{\tiny UTS}}$ for 800--1200 MPa, the range over which dimples are observed on the fracture surface (see also panel c).}
\label{fig:exp}
\end{figure*}

We performed uniaxial tension tests on cylindrical rods made of a Zr$_{\mbox{\tiny 44}}$Ti$_{\mbox{\tiny 11}}$Cu$_{\mbox{\tiny 10}}$Ni$_{\mbox{\tiny 10}}$Be$_{\mbox{\tiny 25}}$ metallic glass ($T_{\rm g}\=625$ K) of diameter $D_0\=1.8$ mm, over a range of temperatures $T$ and strain rates $\dot\epsilon$. The rods were prepared such that the as-cast state is statistically similar between samples (i.e.,~featuring the same fictive temperature~\cite{ketkaew2018mechanical}), see Supplementary Materials for details~\cite{supp}, ensuring that all of the observed variations in the failure dynamics are due to variations in the control parameters $T$ and $\dot\epsilon$, and not different nonequilibrium histories. The rods were clamped in a universal testing machine and maintained at a temperature $T$ (both below and above $T_{\rm g}$, in the range of $T\=529\!-\!659$ K), before being loaded in tension at a strain rate $\dot\epsilon$ (in the range $\dot\epsilon\=0.2\!-\!1.3$ s$^{-1}$), see~\cite{supp} for details.

For any uniaxial test with prescribed $T$ and $\dot\epsilon$, we measured the applied stress $\sigma$ as a function of the strain $\epsilon$. For each stress-strain curve $\sigma(\epsilon)$, we extracted the peak value, i.e., the ultimate tensile strength (UTS), $\sigma_{\mbox{\tiny UTS}}$. In Fig.~\ref{fig:exp}a, we present the observed macroscopic, postmortem  failure patterns as a function of increasing $\sigma_{\mbox{\tiny UTS}}$, independently of whether its variation has been achieved by varying $T$ or $\dot\epsilon$. At small $\sigma_{\mbox{\tiny UTS}}$, corresponding to high $T$ and low $\dot\epsilon$, necking-mediated failure is observed. At large $\sigma_{\mbox{\tiny UTS}}$, corresponding to low $T$ and high $\dot\epsilon$, oblique, shear-banding-mediated failure is observed. In between, a sequence of macroscopic failure patterns is observed, apparently parameterized by $\sigma_{\mbox{\tiny UTS}}$.  

We roughly identify three intermediate failure patterns, sketched on the top row of Fig.~\ref{fig:exp}b, between necking-mediated failure (leftmost) and shear-banding-mediated failure (rightmost). To better describe and understand the sequence of observed macroscopic failure patterns, we polished the postmortem samples that feature a vanishingly small cross section at failure --- characterizing predominantly necking-mediated failure (the two leftmost sketches on the top row of Fig.~\ref{fig:exp}b) --- along the rod's long axis and imaged it using a scanning electron microscope (see~\cite{supp} for details). For the lowest $\sigma_{\mbox{\tiny UTS}}$, where the neck is long, the bulk of the sample is homogeneous (see leftmost image on the bottom row of Fig.~\ref{fig:exp}b), not revealing clear mesoscale structures, hence indicating homogeneous plastic deformation. At somewhat larger $\sigma_{\mbox{\tiny UTS}}$, for which samples still neck (yet the neck is shorter), the bulk of the sample reveals mesoscopic structures in the form of cavities (see the bottom row of Fig.~\ref{fig:exp}b and the inset therein), indicating void formation and coalescence that leads to mesoscopic cavities in the bulk.

\begin{figure*}[t!]
\includegraphics[width = \textwidth]{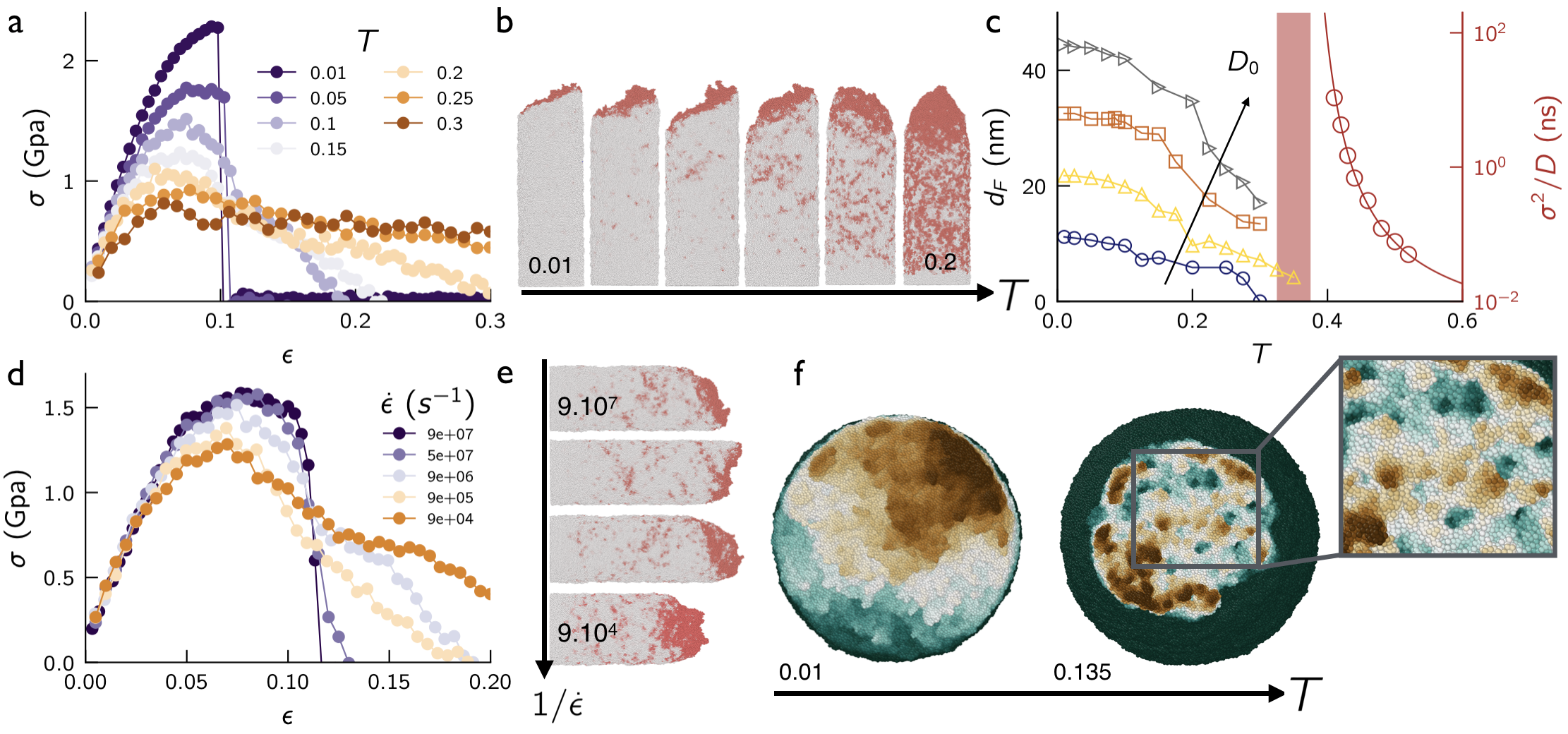}
\caption{\footnotesize (a) $\sigma(\epsilon)$ for various $T$'s and a fixed $\dot{\epsilon}\!=\!9.10^6 s^{-1}$, with $D_0\!=\!11.8$ nm (all in simulational units, see~\cite{supp}). (b) Postmortem global failure patterns for various $T$'s (red particles represent shear-driven plasticity, see~\cite{supp}). (c) Rod's diameter at failure $d_{\rm F}$ (left y-axis) for various $D_0$'s and the relaxation time $\sigma^2/D$ (right y-axis, the solid line is a Vogel-Fulcher-Tammann fit) vs.~$T$, see text for discussion. The red vertical bar provides an estimate for the computer glass transition. (d-e) The same as panels a-b, but for various $\dot\epsilon$'s and a fixed $T\!=\!0.1$. (f) Fractography at two different temperatures ($T\!=\!0.01$ and $0.135$) for $D_0\!=\!23.5$ nm and $\dot{\epsilon}\!=\!5.10^7 s^{-1}$, where the colors represent the depth from green to brown. A zoom in indicates small dimples.}
\label{fig:sim}
\end{figure*}

For postmortem samples that feature a finite cross section at failure, corresponding to the three rightmost sketches on the top row of Fig.~\ref{fig:exp}b, we imaged the fracture surfaces using a scanning electron microscope and performed a fractographic analysis. We identified two distinct fractographic patterns~\cite{spaepen1975fracture,argon1976mechanism,qu2010tensile,qu2013compressive,pan2017ductile,ketkaew2018mechanical,PhysRevLett.94.125510}, veins that are characteristic of localized shear plasticity and dimples that are characteristic of cavities. For intermediate $\sigma_{\mbox{\tiny UTS}}$, necking interrupted by cup-like structures appear in the macroscopic failure pattern (middle sketch on the top row of Fig.~\ref{fig:exp}b), which corresponds to dimples covering the entire fracture surface, see the two images below the middle sketch on the top row of Fig.~\ref{fig:exp}b, where the blue boundary encircles dimples (here the entire surface). For yet higher $\sigma_{\mbox{\tiny UTS}}$, a cup-and-cone-like macroscopic pattern emerges (next-to-rightmost sketch on the top row of Fig.~\ref{fig:exp}b), which corresponds to the coexistence of vein-like patterns at the periphery (``cone'' part, see red encircling lines in the two images below) and dimples at the center (``cup'' part, see blue encircling lines therein). Finally, for shear-banding-mediated failure at high $\sigma_{\mbox{\tiny UTS}}$, shown on the rightmost part of Fig.~\ref{fig:exp}b, vein-like patterns dominate the fracture surface.

The experimental results discussed above indicate that the various observed macroscopic failure modes emerge from the interplay between shear-driven plasticity and dilation-driven void formation/cavitation. These irreversible processes manifest themselves on the fracture surface, respectively, but generally also take place inside the bulk of the glass. To further quantify the fractographic manifestations of these irreversible processes, we present in Fig.~\ref{fig:exp}c the percentage of the fractured area that is covered by dimples (see also panel b) as a function of $\sigma_{\mbox{\tiny UTS}}$, controlled either by varying $\dot\epsilon$ for a fixed $T$ (orange circles) or by varying $T$ at a fixed $\dot\epsilon$ (green squares), see legend. While for small $\sigma_{\mbox{\tiny UTS}}$ the dimple fraction is 100\% by construction (it is simply the tip of the neck, which involves intensive plastic deformation), at higher $\sigma_{\mbox{\tiny UTS}}$ the fraction drops, corresponding to the coexistence of dimples and vein-like patterns, until the latter dominate at the highest $\sigma_{\mbox{\tiny UTS}}$. Interestingly, the two curves appear to overlap, further substantiating the central role played by $\sigma_{\mbox{\tiny UTS}}(\dot\epsilon,T)$. Finally, in Fig.~\ref{fig:exp}d we present the maximal dimple size vs.~$\sigma_{\mbox{\tiny UTS}}\!>\!800$ MPa (on a log-linear scale), revealing a decreasing function.  

The experiments described above span the entire spectrum of tensile failure modes in a glass as a function of a continuous control parameter, and provide insight into the deformation processes involved. Yet, they offer only indirect evidence for the spatiotemporal dynamics that actually determine failure. To start closing this gap, we performed large-scale molecular dynamics simulations of tensile failure of computer glasses, over a wide range of temperatures and strain rates, allowing access to atomistic deformation processes preceding failure inside the glass. We employed computer glasses composed of a 50:50 binary mixture of particles interacting through a modified Lennard-Jones type potential~\cite{richard2021brittle} (see~\cite{supp} for the rationale behind this choice), forming cylindrical rods of length $L_0\=35.3$ nm and various diameters $D_0\=11.8,23.5,35.3$ and $47$ nm (see~\cite{supp} for units conversion rules). The latter correspond to a number of particles $N$ ranging from 300K to 5M. In order to minimize surface effects, we followed the casting procedure of~\cite{shi2010size}. Tensile test simulations were carried out using a massively parallel LAMMPS package~\cite{plimpton1995fast}, and details regarding the tensile loading and thermostating procedures are provided in~\cite{supp}.

While atomistic simulations offer unique and powerful possibilities to resolve the spatiotemporal dynamics on the way to material failure, they are still limited in size and accessible timescales compared to macroscopic glasses (though the employed temperature range is fully consistent with laboratory experiments, including those reported on in Fig.~\ref{fig:exp}). For example, the lowest strain-rate probed in our simulations is about $10^{5}$ s$^{-1}$, significantly larger than in typical experiments. Consequently, our first goal is to understand whether atomistic simulations can recover the sequence of experimental failure modes presented in Fig.~\ref{fig:exp}a, albeit over different length and timescales. In Fig.~\ref{fig:sim}a, we present stress-strain curves $\sigma(\epsilon)$ for computer tensile tests performed on rods of diameter $D_0\=11.8$ nm, over a broad range of $T$'s at a fixed $\dot\epsilon$. The corresponding failure modes are presented in Fig.~\ref{fig:sim}b, where the red regions indicate shear-driven plasticity. The results clearly demonstrate that our atomistic simulations span the entire range of the experimentally observed failure modes, from oblique, shear-banding-mediated failure at low $T$ to necking-mediated failure at higher $T$. These encouraging results give hope that atomistic simulations can offer fundamental insight into the spatiotemporal physics that accompany glass failure.       

At the same time, as stressed above, the computer rods are much smaller than the experimental ones, and it is established that small glass samples exhibit enhanced plasticity compared to their macroscopic counterparts~\cite{guo2007tensile,volkert2008effect,jang2010transition,magagnosc2013tunable,bonfanti2018damage,shi2019size}. Consequently, we expect computer samples to exhibit necking-mediated failure at temperatures that are smaller than the corresponding ones for laboratory samples, when measured relative to the glass temperature $T_{\rm g}$. For the experimental results of Fig.~\ref{fig:exp}a, necking-mediated failure emerges for $T\!>\!T_{\rm g}$. In Fig.~\ref{fig:sim}c (right y-axis), we first estimate $T_{\rm g}$ for the computer samples by plotting the inverse diffusion coefficient (providing a measure of the structural relaxation time) as a function of $T$. By estimating the divergence of the relaxation time (solid line going through the red circles), we estimate the computer glass temperature as $T_{\rm g}\!\simeq\!0.35$ (in simulational units), marked by the red vertical bar.

In Fig.~\ref{fig:sim}c (left y-axis), we also plot the rod's diameter at failure, $d_{\rm F}$, as a function of $T$ for different initial rod's diameters $D_0$. The limit $d_{\rm F}\!\to\!0$ corresponds to the necking limit, i.e.,~to a vanishingly small cross section at failure. For the smallest $D_0$ ($D_0\=11.8$ nm, blue circles, corresponding to the results shown in Fig.~\ref{fig:sim}a-b), the necking limit is reached for $T\!<\!T_{\rm g}$, unlike the laboratory experiments. With increasing $D_0$ (see arrow, different values of $D_0$ correspond to different symbols and colors in Fig.~\ref{fig:sim}c), the necking limit is pushed to higher $T$'s, clearly above $T_{\rm g}$ (while the actual $d_{\rm F}\!\to\!0$ limit is not fully resolved due to computational power constraints), making it consistent with the experimental observations.      

In Fig.~\ref{fig:sim}d-e, we present results indicating that a similar sequence of failure modes (shown in Fig.~\ref{fig:sim}a-b) is observed in our simulations for a fixed $T$ and variable $\dot\epsilon$, in line with the experimental results presented in Fig.~\ref{fig:exp}. In view of this correspondence, we focus on the variation with $T$ hereafter. It is important to note that in the snapshots presented in Figs.~\ref{fig:sim}b,e, we marked shear-driven plasticity (in red), but did not visualize dilation-driven plasticity in the form of void formation (to be considered below). In Fig.~\ref{fig:sim}f, we present the fractography of $D_0\=23.5$ nm rods at two temperatures (one in the predominantly shear-banding-mediated failure regime, left, and the other in the predominantly ``cup'' failure regime, right), see figure caption for details. Small dimples are observed, as highlighted in the zoom in on the right, which are reminiscent of the experimental fractographic observation of Fig.~\ref{fig:exp}b (middle part), though on much smaller lengthscales. 

It is important to note, however, that we do not observe vein-like patterns during shear-banding in our simulations. Veins are commonly attributed to meniscus instabilities caused by a local increase of the temperature and the accompanying reduced viscosity inside the shear band~\cite{argon1976mechanism,qu2010tensile,qu2013compressive}. While we do observe a local temperature rise inside the shear band (both when the system is coupled to a thermostat of a finite relaxation time, as in the simulations reported in this work, and in simulations without thermostating, not reported here), vein-like patterns do not emerge. We suspect that this is the case due to the limited size of our computer samples. Indeed, the heat generated during failure is expected to be proportional to the elastic energy stored (and subsequently released) in the system and thus finite size effects are expected.

\begin{figure*}[t!]
\includegraphics[width = \textwidth]{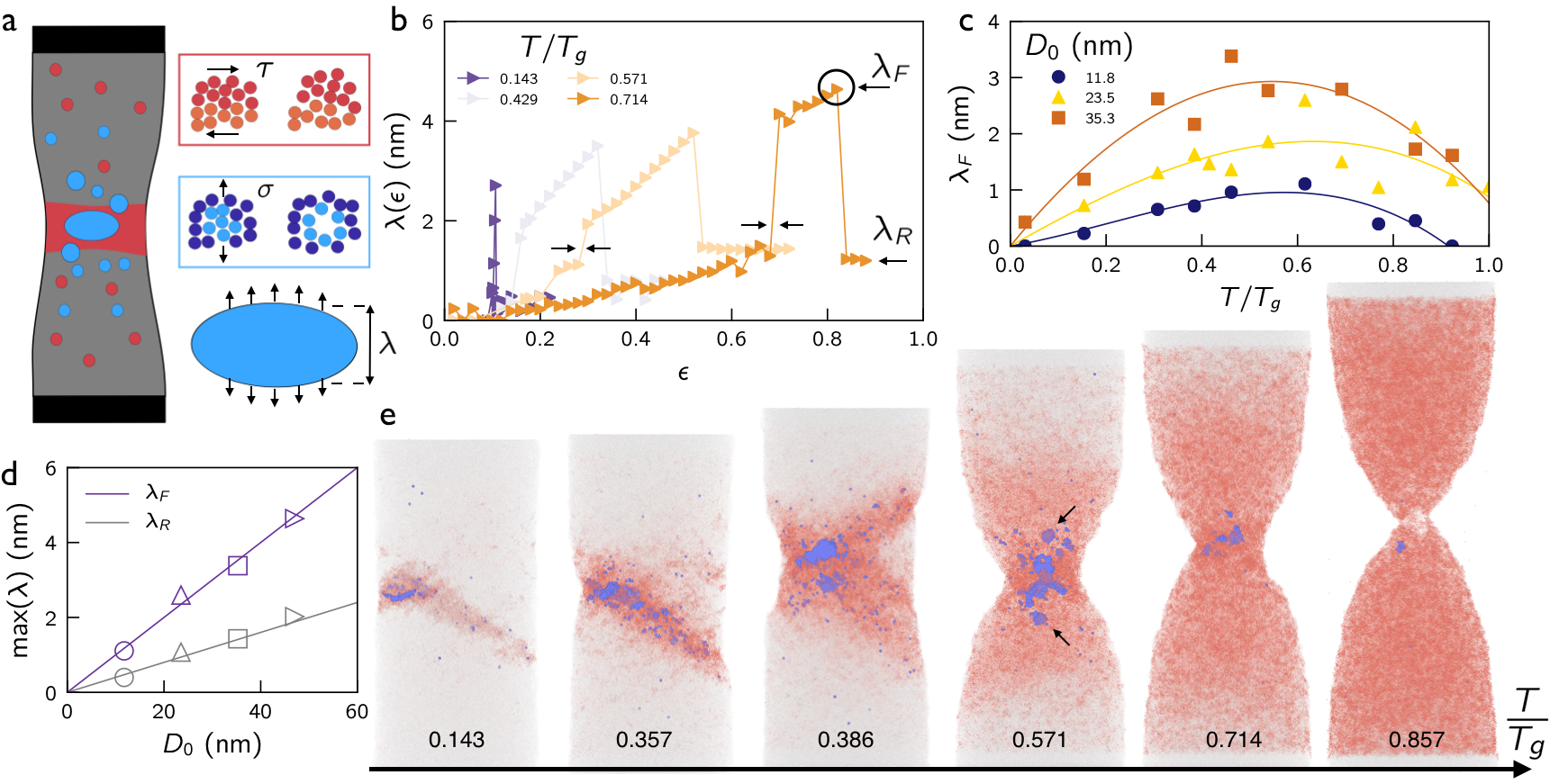}
\caption{\footnotesize (a) A sketch of a deformed rod, where shear-driven plasticity and voids are rendered in red and blue, respectively (see~\cite{supp} for details). (b) The largest cavity size $\lambda$ in the tensile direction vs.~$\epsilon$ for various $T$'s, $\dot{\epsilon}\!=\!5.10^7 s^{-1}$, and $D_0\!=\!47$ nm. See text for discussion, including the meaning of the black circle and various arrows. $\lambda_{\rm F}$ and $\lambda_{\rm R}$ denote the values of $\lambda$ at and after global failure, respectively. (c) $\lambda_F$ vs.~$T/T_{\rm g}$ for various rod's diameters $D_0$ (see legend). See text for extensive discussion. (d) $\lambda_{\rm F}$ and $\lambda_{\rm R}$ vs.~$D_0$ (see legend). The solid lines indicate a linear behavior. (e) Side view snapshots at failure for various $T/T_{\rm g}$ (as indicated), with $D_0\!=\!23.5$ nm. The arrows in the $T/T_{\rm g}\!=\!0.571$ case indicate residual cavities inside the bulk.}
\label{fig:stz_ttz}
\end{figure*}

Our next goal is to investigate the spatiotemporal interplay between shear-driven plasticity and dilation-driven void formation inside the glass prior to --- and approaching --- material failure. In Fig.~\ref{fig:stz_ttz}a, we present a sketch of a rod under tension, illustrating the coexistence of shear-driven plastic events (red) and dilation-driven void formation (blue), see also sketches in the rectangles on the right. As material failure is accompanied by decohesion and the creation of internal free surfaces that cannot sustain stress, we focus on the largest cavity (a cluster of coalesced voids) as a representative indicator of material decohesion under tension, in particular on its size $\lambda$ in the tensile direction (see bottom-right panel of Fig.~\ref{fig:stz_ttz}a). Shear-driven plasticity is monitored based on the best fit of the non-affine displacement using the common $D^2_{\rm min}$ field defined in~\cite{falk1998dynamics}. We detect the nucleation of voids by inserting ghost particles such as done e.g.~in.~\cite{menzl2016molecular,galimzyanov2022cavity}. Details are provided in~\cite{supp}.

In Fig.~\ref{fig:stz_ttz}b, we present the evolution of $\lambda$ with strain $\epsilon$ for various $T$'s. Focusing first on the largest $T$ presented (corresponding to the next-to-rightmost sample in Fig.~\ref{fig:stz_ttz}e), we observe that $\lambda$ increases with $\epsilon$ rather smoothly until a large, discrete/discontinuous jump takes place (marked by the two arrows). This jump corresponds to the coalescence of two (or more) mesoscopic cavities (see~\cite{supp}), yet it does not lead to global failure. Instead, $\lambda$ continues to increase rather smoothly until global failure occurs, defined by $\lambda\=\lambda_{\rm F}$ (marked on the figure by a black circle). We associate the smooth increase in $\lambda$ with void growth facilitated by shear-driven plasticity, i.e., the plastic expansion of a single cavity within the glass~\cite{bouchbinder2007stability,tang2020cavitation}. Interestingly, after failure, cavities of finite size $\lambda_{\rm R}$ remain locked-in inside the glass, exactly as observed experimentally in Fig.~\ref{fig:exp}b (next-to-leftmost panel). As $T$ is decreased, $\lambda(\epsilon)$ features similar properties, yet it is pushed to lower strains and smaller values of $\lambda_{\rm F}$. At the lowest $T$, failure becomes very abrupt, associated with a rapid (in strain $\epsilon$) formation of a system-spanning (in the diameter direction) cavity, essentially a crack.  

In Fig.~\ref{fig:stz_ttz}c, we focus on $\lambda_{\rm F}$ as an important quantifier of the overall failure process. In particular, $\lambda_{\rm F}$ is plotted as a function of $T$, for different $D_0$'s. For all $D_0$'s, $\lambda_{\rm F}(T)$ is nonmonotonic, featuring low values of $\lambda_{\rm F}$ for both low and high $T$ (approaching $T_{\rm g}$), reaching a maximum at an intermediate $T$. To shed additional light on this result, we present in Fig.~\ref{fig:stz_ttz}e snapshots of samples deformed at different $T$'s, at their failure strain (i.e.,~corresponding to the strain defining $\lambda_{\rm F}$ in panel b). In these snapshots (and in the tensile test simulations leading to them), we spatiotemporally tracked shear-driven plasticity (red) and dilation-driven void formation (blue) at the particle level (as explained in~\cite{supp}), to gain insight into their  collective evolution and mutual interactions. 

At all $T$'s, shear-driven plasticity takes place first and dilation-driven void formation follows in spatial locations where shear-driven plasticity already took place, suggesting a causal relation between the two basic processes. These observations indicate that shear-driven plasticity leads to the softening of glassy structures~\cite{lewandowski2006temperature,mota2021enhancing,liu2017shear}, which in turn apparently reduces typical barriers for void formation~\cite{liu2017shear}. The microscopic mechanism for this coupling between shear-driven plasticity and void formation might be the (transient or persistent) creation of free-volume. That is, shear-driven plasticity is known to be accompanied by free-volume creation (see, for example, the recent experiments on colloidal glasses~\cite{lu2018dilatancy,wang2022elastic}, which echo similar observations in metallic glasses~\cite{klaumunzer2011probing}) and regions of higher free-volume might be more susceptible to void formation.

Yet, free-volume creation should be distinguished from void formation, and in particular from cavitation, as the latter involves the formation of new surfaces within the glass (i.e.~it involves surface energies) and is influenced by the hydrostatic tension (i.e.,~the trace of the stress tensor)~\cite{murali2011atomic,rycroft2012fracture,guan2013cavitation,vasoya2016notch}. The hydrostatic tension, in turn, is also reduced through stress relaxation mediated by shear-driven plasticity. Yet another factor at play is the geometric reduction in the cross section of the rod, driven by shear-driven plasticity (which gives rise to the development of a neck at relatively high $T$), leading to an increase in the local tensile stress. With these interrelated and intrinsically coupled physical processes in mind, we now aim at rationalizing the main observations in Fig.~\ref{fig:stz_ttz}c,e.

At low $T$, shear-driven plasticity is localized into a narrow shear-band, without any appreciable reduction in the rod's cross section (cf.~Fig.~\ref{fig:sim}c), and without significant stress relaxation, leading to high tensile stresses (cf.~Fig.~\ref{fig:sim}a). Consequently, voids form inside the shear-band and rapidly transform into a catastrophic crack under the high tension~\cite{falk1999molecular,murali2013shear,luo2015tensile,shao2014direct}, without significantly growing in the tensile direction, hence $\lambda_{\rm F}$ is small in this low $T$ regime (cf.~Fig.~\ref{fig:stz_ttz}c). Note, however, that the number of independent cavities grows with the volume of the shear band and thus scales with $D_0^2$ (see~\cite{supp}). We find that the cluster distributions are consistent with the picture provided by homogeneous nucleation theory, with cavitation barriers of only a few $k_{\rm B}T$ (see~\cite{supp}), as already pointed out in~\cite{wright2003free}.

With growing $T$, shear-driven plasticity is more diffused. Moreover, it evolves over larger strain intervals and leads to the reduction in the tensile stress, allowing voids to nucleate, grow and coalesce. This leads to larger cavities at failure, hence $\lambda_{\rm F}$ increases with $T$ in Fig.~\ref{fig:stz_ttz}c. The larger $\lambda_{\rm F}$ values manifest themselves as ``cup'' structures on the fracture surface, see Fig.~\ref{fig:sim}b,f for the relevant simulational results and Fig.~\ref{fig:exp}b for the corresponding experimental observations. The larger $\lambda_{\rm F}$ values are also accompanied, in this $T$ regime, by residual cavities within the glass, cf.~the finite value of $\lambda_{\rm R}$ in Fig.~\ref{fig:stz_ttz}b and the arrows in Fig.~\ref{fig:stz_ttz}e. Finally, while shear-driven plasticity is more diffused in this $T$ regime, shear localization still takes place (possibly along two major orientations, cf.~Fig.~\ref{fig:stz_ttz}e with $T/T_{\rm g}\!=\!0.386$), leading to a ``cone'' structure close to the periphery of the failed samples. 

As $T$ is further increased, shear-driven plasticity becomes spatially extended, leading to a reduction in both the overall tensile stress and in the rod's local cross section as global failure is approached. The reduction in the tensile stress leads to decreasing values of $\lambda_{\rm F}$, making $\lambda_{\rm F}(T)$ nonmonotonic. The geometric reduction of the cross section as a neck is formed, apparently maintains the local tensile stress sufficiently large to keep $\lambda_{\rm F}$ finite. The spatially extended, even more diffused, nature of shear-driven plasticity in this regime also leads to the disappearance of the ``cone'' structure. As $T$ is further increased, stress relaxation by shear-driven plasticity accompanying necking is so efficient, that cavities cannot grow anymore and $\lambda_{\rm F}$ becomes vanishingly small, consistent with the absence of bulk cavities in the experimental data shown on the leftmost part of Fig.~\ref{fig:exp}b. 

The nonmonotonic behavior of $\lambda_{\rm F}(T)$ is fully consistent with the decrease in the dimple area fraction with increasing $\sigma_{\mbox{\tiny UTS}}$ --- corresponding to decreasing $T$ --- in the experimental data of Fig.~\ref{fig:exp}c and with the decrease in the maximal dimple size with increasing $\sigma_{\mbox{\tiny UTS}}$ in Fig.~\ref{fig:exp}d. In fact, it predicts that fractographic measurements of the maximal dimple size for $\sigma_{\mbox{\tiny UTS}}$ below 800 MPa will reveal a peak, before dropping. To further connect our simulational findings to the experimental data, which feature significantly larger lengthscales as discussed above, we present 
in Fig.~\ref{fig:stz_ttz}d both $\lambda_{\rm F}$ and $\lambda_{\rm R}$ as a function of the rod's diameter $D_0$. Both quantities follow a linear relation with $D_0$, where for the former we have $\lambda_{\rm F}\!\simeq\!D_0/10$. Extrapolating this result to the experimental lengthscale, where $D_0$ is in the mm range, we expect $\lambda_{\rm F}$ to be in the $10^{-1}$ mm range, in the right ballpark of the data presented in Fig.~\ref{fig:exp}d.    

In summary, by combining extensive experiments on metallic glasses and large-scale molecular dynamic simulations of computer glasses, we provided physical insight into the transition in tensile failure modes of glasses, from necking-mediated failure to shear-banding-mediated failure, as a function of the applied strain-rate $\dot\epsilon$ and the temperature $T$. At the macroscopic level, we showed that the sequence of global failure modes depends in a unified manner on the ultimate tensile stress $\sigma_{\mbox{\tiny UTS}}(\dot\epsilon,T)$. At the mesoscopic level, using experimental fractography and postmortem sample polishing, as well as particle-scale quantification of computer simulations, we showed that the interplay between shear-driven plasticity and dilation-driven void formation controls the global failure mode.  

More specifically, we showed that the spatiotemporal evolution of shear-driven plasticity and dilation-driven void formation --- and their intrinsic coupling in space and time --- account for the temperature and strain-rate dependence of tensile failure modes in glasses. Our large-scale computer simulations demonstrate that the above mentioned interplay can be quantified through the size of the largest cavity (a cluster of coalesced voids) in the tensile direction, which exhibits a nonmonotonic temperature dependence. Furthermore, we show that the latter scales linearly with the glass sample's diameter, which upon extrapolation offers a possible way to bridge the vast difference in lengthscales between computer and laboratory glasses.

Our results also pose various questions for future investigations. Among these, we would like to highlight the need to determine whether the transition between the various macroscopic failure modes as a function of temperature and strain-rate is continuous or discontinuous, and the need to understand the effect of the initial nonequilibrium glass state (fictive temperature), which was kept fixed in this study, on the failure mode. Finally, our results highlight the pressing need to develop elasto-plastic constitutive models of glassy deformation, which self-consistently account for both shear- and dilation-driven spatiotemporal, dissipative dynamics.\\

{\em Acknowledgments.} D.R.~acknowledges support by the H2020-MSCA-IF-2020 project ToughMG  (No.~101024057). This work was performed using HPC resources from GENCI-IDRIS (Grant 2022-AD010913428). E.B.~acknowledges support from the Ben May Center for Chemical Theory and Computation and the Harold Perlman Family. E.L.~acknowledges support by the National Science Foundation Graduate Research Fellowship under Grant No.~2139841. The experimental work was supported by the Office of Naval Research under grant No.~N00014-20-1-2200.


\begin{thebibliography}{51}%
\makeatletter
\providecommand \@ifxundefined [1]{%
 \@ifx{#1\undefined}
}%
\providecommand \@ifnum [1]{%
 \ifnum #1\expandafter \@firstoftwo
 \else \expandafter \@secondoftwo
 \fi
}%
\providecommand \@ifx [1]{%
 \ifx #1\expandafter \@firstoftwo
 \else \expandafter \@secondoftwo
 \fi
}%
\providecommand \natexlab [1]{#1}%
\providecommand \enquote  [1]{``#1''}%
\providecommand \bibnamefont  [1]{#1}%
\providecommand \bibfnamefont [1]{#1}%
\providecommand \citenamefont [1]{#1}%
\providecommand \href@noop [0]{\@secondoftwo}%
\providecommand \href [0]{\begingroup \@sanitize@url \@href}%
\providecommand \@href[1]{\@@startlink{#1}\@@href}%
\providecommand \@@href[1]{\endgroup#1\@@endlink}%
\providecommand \@sanitize@url [0]{\catcode `\\12\catcode `\$12\catcode
  `\&12\catcode `\#12\catcode `\^12\catcode `\_12\catcode `\%12\relax}%
\providecommand \@@startlink[1]{}%
\providecommand \@@endlink[0]{}%
\providecommand \url  [0]{\begingroup\@sanitize@url \@url }%
\providecommand \@url [1]{\endgroup\@href {#1}{\urlprefix }}%
\providecommand \urlprefix  [0]{URL }%
\providecommand \Eprint [0]{\href }%
\providecommand \doibase [0]{https://doi.org/}%
\providecommand \selectlanguage [0]{\@gobble}%
\providecommand \bibinfo  [0]{\@secondoftwo}%
\providecommand \bibfield  [0]{\@secondoftwo}%
\providecommand \translation [1]{[#1]}%
\providecommand \BibitemOpen [0]{}%
\providecommand \bibitemStop [0]{}%
\providecommand \bibitemNoStop [0]{.\EOS\space}%
\providecommand \EOS [0]{\spacefactor3000\relax}%
\providecommand \BibitemShut  [1]{\csname bibitem#1\endcsname}%
\let\auto@bib@innerbib\@empty
\bibitem [{\citenamefont {Spaepen}(1977)}]{spaepen1977microscopic}%
  \BibitemOpen
  \bibfield  {author} {\bibinfo {author} {\bibfnamefont {F.}~\bibnamefont
  {Spaepen}},\ }\bibfield  {title} {\bibinfo {title} {A microscopic mechanism
  for steady state inhomogeneous flow in metallic glasses},\ }\href
  {https://doi.org/https://doi.org/10.1016/0001-6160(77)90232-2} {\bibfield
  {journal} {\bibinfo  {journal} {Acta metallurgica}\ }\textbf {\bibinfo
  {volume} {25}},\ \bibinfo {pages} {407} (\bibinfo {year} {1977})}\BibitemShut
  {NoStop}%
\bibitem [{\citenamefont {Lu}\ \emph {et~al.}(2003)\citenamefont {Lu},
  \citenamefont {Ravichandran},\ and\ \citenamefont
  {Johnson}}]{lu2003deformation}%
  \BibitemOpen
  \bibfield  {author} {\bibinfo {author} {\bibfnamefont {J.}~\bibnamefont
  {Lu}}, \bibinfo {author} {\bibfnamefont {G.}~\bibnamefont {Ravichandran}},\
  and\ \bibinfo {author} {\bibfnamefont {W.~L.}\ \bibnamefont {Johnson}},\
  }\bibfield  {title} {\bibinfo {title} {Deformation behavior of the zr41.
  2ti13. 8cu12. 5ni10be22. 5 bulk metallic glass over a wide range of
  strain-rates and temperatures},\ }\href
  {https://doi.org/https://doi.org/10.1016/S1359-6454(03)00164-2} {\bibfield
  {journal} {\bibinfo  {journal} {Acta materialia}\ }\textbf {\bibinfo {volume}
  {51}},\ \bibinfo {pages} {3429} (\bibinfo {year} {2003})}\BibitemShut
  {NoStop}%
\bibitem [{\citenamefont {Wang}\ \emph {et~al.}(2005)\citenamefont {Wang},
  \citenamefont {Shen}, \citenamefont {Sun}, \citenamefont {Lu}, \citenamefont
  {Stachurski},\ and\ \citenamefont {Zhou}}]{wang2005tensile}%
  \BibitemOpen
  \bibfield  {author} {\bibinfo {author} {\bibfnamefont {G.}~\bibnamefont
  {Wang}}, \bibinfo {author} {\bibfnamefont {J.}~\bibnamefont {Shen}}, \bibinfo
  {author} {\bibfnamefont {J.}~\bibnamefont {Sun}}, \bibinfo {author}
  {\bibfnamefont {Z.}~\bibnamefont {Lu}}, \bibinfo {author} {\bibfnamefont
  {Z.}~\bibnamefont {Stachurski}},\ and\ \bibinfo {author} {\bibfnamefont
  {B.}~\bibnamefont {Zhou}},\ }\bibfield  {title} {\bibinfo {title} {Tensile
  fracture characteristics and deformation behavior of a zr-based bulk metallic
  glass at high temperatures},\ }\href
  {https://doi.org/https://doi.org/10.1016/j.intermet.2004.10.011} {\bibfield
  {journal} {\bibinfo  {journal} {Intermetallics}\ }\textbf {\bibinfo {volume}
  {13}},\ \bibinfo {pages} {642} (\bibinfo {year} {2005})}\BibitemShut
  {NoStop}%
\bibitem [{\citenamefont {Vormelker}\ \emph {et~al.}(2008)\citenamefont
  {Vormelker}, \citenamefont {Vatamanu}, \citenamefont {Kecskes},\ and\
  \citenamefont {Lewandowski}}]{vormelker2008effects}%
  \BibitemOpen
  \bibfield  {author} {\bibinfo {author} {\bibfnamefont {A.~H.}\ \bibnamefont
  {Vormelker}}, \bibinfo {author} {\bibfnamefont {O.}~\bibnamefont {Vatamanu}},
  \bibinfo {author} {\bibfnamefont {L.}~\bibnamefont {Kecskes}},\ and\ \bibinfo
  {author} {\bibfnamefont {J.}~\bibnamefont {Lewandowski}},\ }\bibfield
  {title} {\bibinfo {title} {Effects of test temperature and loading conditions
  on the tensile properties of a zr-based bulk metallic glass},\ }\href
  {https://doi.org/https://doi.org/10.1007/s11661-007-9410-4} {\bibfield
  {journal} {\bibinfo  {journal} {Metall. Mater. Trans. A}\ }\textbf {\bibinfo
  {volume} {39}},\ \bibinfo {pages} {1922} (\bibinfo {year}
  {2008})}\BibitemShut {NoStop}%
\bibitem [{\citenamefont {Schroers}\ and\ \citenamefont
  {Johnson}(2004)}]{schroers2004ductile}%
  \BibitemOpen
  \bibfield  {author} {\bibinfo {author} {\bibfnamefont {J.}~\bibnamefont
  {Schroers}}\ and\ \bibinfo {author} {\bibfnamefont {W.~L.}\ \bibnamefont
  {Johnson}},\ }\bibfield  {title} {\bibinfo {title} {Ductile bulk metallic
  glass},\ }\href
  {https://doi.org/https://doi.org/10.1103/PhysRevLett.93.255506} {\bibfield
  {journal} {\bibinfo  {journal} {Phys. Rev. Lett.}\ }\textbf {\bibinfo
  {volume} {93}},\ \bibinfo {pages} {255506} (\bibinfo {year}
  {2004})}\BibitemShut {NoStop}%
\bibitem [{\citenamefont {Bouchbinder}\ \emph
  {et~al.}(2008{\natexlab{a}})\citenamefont {Bouchbinder}, \citenamefont {Lo},
  \citenamefont {Procaccia},\ and\ \citenamefont
  {Shtilerman}}]{bouchbinder2007stability}%
  \BibitemOpen
  \bibfield  {author} {\bibinfo {author} {\bibfnamefont {E.}~\bibnamefont
  {Bouchbinder}}, \bibinfo {author} {\bibfnamefont {T.-S.}\ \bibnamefont {Lo}},
  \bibinfo {author} {\bibfnamefont {I.}~\bibnamefont {Procaccia}},\ and\
  \bibinfo {author} {\bibfnamefont {E.}~\bibnamefont {Shtilerman}},\ }\bibfield
   {title} {\bibinfo {title} {Stability of an expanding circular cavity and the
  failure of amorphous solids},\ }\href
  {https://doi.org/10.1103/PhysRevE.78.026124} {\bibfield  {journal} {\bibinfo
  {journal} {Phys. Rev. E}\ }\textbf {\bibinfo {volume} {78}},\ \bibinfo
  {pages} {026124} (\bibinfo {year} {2008}{\natexlab{a}})}\BibitemShut
  {NoStop}%
\bibitem [{\citenamefont {Bouchaud}\ \emph {et~al.}(2008)\citenamefont
  {Bouchaud}, \citenamefont {Boivin}, \citenamefont {Pouchou}, \citenamefont
  {Bonamy}, \citenamefont {Poon},\ and\ \citenamefont
  {Ravichandran}}]{bouchaud2008fracture}%
  \BibitemOpen
  \bibfield  {author} {\bibinfo {author} {\bibfnamefont {E.}~\bibnamefont
  {Bouchaud}}, \bibinfo {author} {\bibfnamefont {D.}~\bibnamefont {Boivin}},
  \bibinfo {author} {\bibfnamefont {J.-L.}\ \bibnamefont {Pouchou}}, \bibinfo
  {author} {\bibfnamefont {D.}~\bibnamefont {Bonamy}}, \bibinfo {author}
  {\bibfnamefont {B.}~\bibnamefont {Poon}},\ and\ \bibinfo {author}
  {\bibfnamefont {G.}~\bibnamefont {Ravichandran}},\ }\bibfield  {title}
  {\bibinfo {title} {Fracture through cavitation in a metallic glass},\ }\href
  {https://doi.org/10.1209/0295-5075/83/66006} {\bibfield  {journal} {\bibinfo
  {journal} {EPL}\ }\textbf {\bibinfo {volume} {83}},\ \bibinfo {pages} {66006}
  (\bibinfo {year} {2008})}\BibitemShut {NoStop}%
\bibitem [{\citenamefont {Jiang}\ \emph {et~al.}(2008)\citenamefont {Jiang},
  \citenamefont {Ling}, \citenamefont {Meng},\ and\ \citenamefont
  {Dai}}]{jiang2008energy}%
  \BibitemOpen
  \bibfield  {author} {\bibinfo {author} {\bibfnamefont {M.}~\bibnamefont
  {Jiang}}, \bibinfo {author} {\bibfnamefont {Z.}~\bibnamefont {Ling}},
  \bibinfo {author} {\bibfnamefont {J.}~\bibnamefont {Meng}},\ and\ \bibinfo
  {author} {\bibfnamefont {L.}~\bibnamefont {Dai}},\ }\bibfield  {title}
  {\bibinfo {title} {Energy dissipation in fracture of bulk metallic glasses
  via inherent competition between local softening and quasi-cleavage},\ }\href
  {https://doi.org/https://doi.org/10.1080/14786430701864753} {\bibfield
  {journal} {\bibinfo  {journal} {Philosophical Magazine}\ }\textbf {\bibinfo
  {volume} {88}},\ \bibinfo {pages} {407} (\bibinfo {year} {2008})}\BibitemShut
  {NoStop}%
\bibitem [{\citenamefont {Bouchbinder}\ \emph
  {et~al.}(2008{\natexlab{b}})\citenamefont {Bouchbinder}, \citenamefont {Lo},\
  and\ \citenamefont {Procaccia}}]{bouchbinder2008dynamic}%
  \BibitemOpen
  \bibfield  {author} {\bibinfo {author} {\bibfnamefont {E.}~\bibnamefont
  {Bouchbinder}}, \bibinfo {author} {\bibfnamefont {T.-S.}\ \bibnamefont
  {Lo}},\ and\ \bibinfo {author} {\bibfnamefont {I.}~\bibnamefont
  {Procaccia}},\ }\bibfield  {title} {\bibinfo {title} {Dynamic failure in
  amorphous solids via a cavitation instability},\ }\href
  {https://doi.org/10.1103/PhysRevE.77.025101} {\bibfield  {journal} {\bibinfo
  {journal} {Phys. Rev. E}\ }\textbf {\bibinfo {volume} {77}},\ \bibinfo
  {pages} {025101} (\bibinfo {year} {2008}{\natexlab{b}})}\BibitemShut
  {NoStop}%
\bibitem [{\citenamefont {Ketkaew}\ \emph {et~al.}(2018)\citenamefont
  {Ketkaew}, \citenamefont {Chen}, \citenamefont {Wang}, \citenamefont {Datye},
  \citenamefont {Fan}, \citenamefont {Pereira}, \citenamefont {Schwarz},
  \citenamefont {Liu}, \citenamefont {Yamada}, \citenamefont {Dmowski} \emph
  {et~al.}}]{ketkaew2018mechanical}%
  \BibitemOpen
  \bibfield  {author} {\bibinfo {author} {\bibfnamefont {J.}~\bibnamefont
  {Ketkaew}}, \bibinfo {author} {\bibfnamefont {W.}~\bibnamefont {Chen}},
  \bibinfo {author} {\bibfnamefont {H.}~\bibnamefont {Wang}}, \bibinfo {author}
  {\bibfnamefont {A.}~\bibnamefont {Datye}}, \bibinfo {author} {\bibfnamefont
  {M.}~\bibnamefont {Fan}}, \bibinfo {author} {\bibfnamefont {G.}~\bibnamefont
  {Pereira}}, \bibinfo {author} {\bibfnamefont {U.~D.}\ \bibnamefont
  {Schwarz}}, \bibinfo {author} {\bibfnamefont {Z.}~\bibnamefont {Liu}},
  \bibinfo {author} {\bibfnamefont {R.}~\bibnamefont {Yamada}}, \bibinfo
  {author} {\bibfnamefont {W.}~\bibnamefont {Dmowski}}, \emph {et~al.},\
  }\bibfield  {title} {\bibinfo {title} {Mechanical glass transition revealed
  by the fracture toughness of metallic glasses},\ }\href
  {https://doi.org/https://doi.org/10.1038/s41467-018-05682-8} {\bibfield
  {journal} {\bibinfo  {journal} {Nat. Commun.}\ }\textbf {\bibinfo {volume}
  {9}},\ \bibinfo {pages} {1} (\bibinfo {year} {2018})}\BibitemShut {NoStop}%
\bibitem [{sup()}]{supp}%
  \BibitemOpen
  \href@noop {} {}\bibinfo {howpublished} {See Supplementary Materials at
  \url{DOI}, which include Refs.
  \cite{ketkaew2018mechanical,richard2021brittle,shi2010size,lerner2019mechanical,bonfanti2018damage,plimpton1995fast,falk1998dynamics,shi2016creating,zhao2023fracture,he2019critical,luo2015tensile,tang2022crack,murali2011atomic,puaduraru2010computer,wright2003free}}\BibitemShut
  {NoStop}%
\bibitem [{\citenamefont {Spaepen}(1975)}]{spaepen1975fracture}%
  \BibitemOpen
  \bibfield  {author} {\bibinfo {author} {\bibfnamefont {F.}~\bibnamefont
  {Spaepen}},\ }\bibfield  {title} {\bibinfo {title} {On the fracture
  morphology of metallic glasses},\ }\href
  {https://doi.org/https://doi.org/10.1016/0001-6160(75)90102-9} {\bibfield
  {journal} {\bibinfo  {journal} {Acta Metallurgica}\ }\textbf {\bibinfo
  {volume} {23}},\ \bibinfo {pages} {615} (\bibinfo {year} {1975})}\BibitemShut
  {NoStop}%
\bibitem [{\citenamefont {Argon}\ and\ \citenamefont
  {Salama}(1976)}]{argon1976mechanism}%
  \BibitemOpen
  \bibfield  {author} {\bibinfo {author} {\bibfnamefont {A.}~\bibnamefont
  {Argon}}\ and\ \bibinfo {author} {\bibfnamefont {M.}~\bibnamefont {Salama}},\
  }\bibfield  {title} {\bibinfo {title} {The mechanism of fracture in glassy
  materials capable of some inelastic deformation},\ }\href
  {https://doi.org/https://doi.org/10.1016/0025-5416(76)90198-1} {\bibfield
  {journal} {\bibinfo  {journal} {Mater. Sci. Eng.}\ }\textbf {\bibinfo
  {volume} {23}},\ \bibinfo {pages} {219} (\bibinfo {year} {1976})}\BibitemShut
  {NoStop}%
\bibitem [{\citenamefont {Qu}\ \emph {et~al.}(2010)\citenamefont {Qu},
  \citenamefont {Stoica}, \citenamefont {Eckert},\ and\ \citenamefont
  {Zhang}}]{qu2010tensile}%
  \BibitemOpen
  \bibfield  {author} {\bibinfo {author} {\bibfnamefont {R.}~\bibnamefont
  {Qu}}, \bibinfo {author} {\bibfnamefont {M.}~\bibnamefont {Stoica}}, \bibinfo
  {author} {\bibfnamefont {J.}~\bibnamefont {Eckert}},\ and\ \bibinfo {author}
  {\bibfnamefont {Z.}~\bibnamefont {Zhang}},\ }\bibfield  {title} {\bibinfo
  {title} {Tensile fracture morphologies of bulk metallic glass},\ }\href
  {https://doi.org/https://doi.org/10.1063/1.3487968} {\bibfield  {journal}
  {\bibinfo  {journal} {J. Appl. Phys.}\ }\textbf {\bibinfo {volume} {108}},\
  \bibinfo {pages} {063509} (\bibinfo {year} {2010})}\BibitemShut {NoStop}%
\bibitem [{\citenamefont {Qu}\ and\ \citenamefont
  {Zhang}(2013)}]{qu2013compressive}%
  \BibitemOpen
  \bibfield  {author} {\bibinfo {author} {\bibfnamefont {R.}~\bibnamefont
  {Qu}}\ and\ \bibinfo {author} {\bibfnamefont {Z.}~\bibnamefont {Zhang}},\
  }\bibfield  {title} {\bibinfo {title} {Compressive fracture morphology and
  mechanism of metallic glass},\ }\href
  {https://doi.org/https://doi.org/10.1063/1.4830029} {\bibfield  {journal}
  {\bibinfo  {journal} {J. Appl. Phys.}\ }\textbf {\bibinfo {volume} {114}},\
  \bibinfo {pages} {193504} (\bibinfo {year} {2013})}\BibitemShut {NoStop}%
\bibitem [{\citenamefont {Pan}\ \emph {et~al.}(2017)\citenamefont {Pan},
  \citenamefont {Wang},\ and\ \citenamefont {Li}}]{pan2017ductile}%
  \BibitemOpen
  \bibfield  {author} {\bibinfo {author} {\bibfnamefont {J.}~\bibnamefont
  {Pan}}, \bibinfo {author} {\bibfnamefont {Y.}~\bibnamefont {Wang}},\ and\
  \bibinfo {author} {\bibfnamefont {Y.}~\bibnamefont {Li}},\ }\bibfield
  {title} {\bibinfo {title} {Ductile fracture in notched bulk metallic
  glasses},\ }\href
  {https://doi.org/https://doi.org/10.1016/j.actamat.2017.06.048} {\bibfield
  {journal} {\bibinfo  {journal} {Acta Materialia}\ }\textbf {\bibinfo {volume}
  {136}},\ \bibinfo {pages} {126} (\bibinfo {year} {2017})}\BibitemShut
  {NoStop}%
\bibitem [{\citenamefont {Xi}\ \emph {et~al.}(2005)\citenamefont {Xi},
  \citenamefont {Zhao}, \citenamefont {Pan}, \citenamefont {Wang},
  \citenamefont {Wu},\ and\ \citenamefont
  {Lewandowski}}]{PhysRevLett.94.125510}%
  \BibitemOpen
  \bibfield  {author} {\bibinfo {author} {\bibfnamefont {X.~K.}\ \bibnamefont
  {Xi}}, \bibinfo {author} {\bibfnamefont {D.~Q.}\ \bibnamefont {Zhao}},
  \bibinfo {author} {\bibfnamefont {M.~X.}\ \bibnamefont {Pan}}, \bibinfo
  {author} {\bibfnamefont {W.~H.}\ \bibnamefont {Wang}}, \bibinfo {author}
  {\bibfnamefont {Y.}~\bibnamefont {Wu}},\ and\ \bibinfo {author}
  {\bibfnamefont {J.~J.}\ \bibnamefont {Lewandowski}},\ }\bibfield  {title}
  {\bibinfo {title} {Fracture of brittle metallic glasses: Brittleness or
  plasticity},\ }\href {https://doi.org/10.1103/PhysRevLett.94.125510}
  {\bibfield  {journal} {\bibinfo  {journal} {Phys. Rev. Lett.}\ }\textbf
  {\bibinfo {volume} {94}},\ \bibinfo {pages} {125510} (\bibinfo {year}
  {2005})}\BibitemShut {NoStop}%
\bibitem [{\citenamefont {Richard}\ \emph {et~al.}(2021)\citenamefont
  {Richard}, \citenamefont {Lerner},\ and\ \citenamefont
  {Bouchbinder}}]{richard2021brittle}%
  \BibitemOpen
  \bibfield  {author} {\bibinfo {author} {\bibfnamefont {D.}~\bibnamefont
  {Richard}}, \bibinfo {author} {\bibfnamefont {E.}~\bibnamefont {Lerner}},\
  and\ \bibinfo {author} {\bibfnamefont {E.}~\bibnamefont {Bouchbinder}},\
  }\bibfield  {title} {\bibinfo {title} {Brittle-to-ductile transitions in
  glasses: Roles of soft defects and loading geometry},\ }\href
  {https://doi.org/https://doi.org/10.1557/s43577-021-00171-8} {\bibfield
  {journal} {\bibinfo  {journal} {MRS Bulletin}\ }\textbf {\bibinfo {volume}
  {46}},\ \bibinfo {pages} {902} (\bibinfo {year} {2021})}\BibitemShut
  {NoStop}%
\bibitem [{\citenamefont {Shi}(2010)}]{shi2010size}%
  \BibitemOpen
  \bibfield  {author} {\bibinfo {author} {\bibfnamefont {Y.}~\bibnamefont
  {Shi}},\ }\bibfield  {title} {\bibinfo {title} {Size-independent shear band
  formation in amorphous nanowires made from simulated casting},\ }\href
  {https://doi.org/https://doi.org/10.1063/1.3340908} {\bibfield  {journal}
  {\bibinfo  {journal} {Appl. Phys. Lett.}\ }\textbf {\bibinfo {volume} {96}},\
  \bibinfo {pages} {121909} (\bibinfo {year} {2010})}\BibitemShut {NoStop}%
\bibitem [{\citenamefont {Plimpton}(1995)}]{plimpton1995fast}%
  \BibitemOpen
  \bibfield  {author} {\bibinfo {author} {\bibfnamefont {S.}~\bibnamefont
  {Plimpton}},\ }\bibfield  {title} {\bibinfo {title} {Fast parallel algorithms
  for short-range molecular dynamics},\ }\href
  {https://doi.org/https://doi.org/10.1006/jcph.1995.1039} {\bibfield
  {journal} {\bibinfo  {journal} {J. Comput. Phys.}\ }\textbf {\bibinfo
  {volume} {117}},\ \bibinfo {pages} {1} (\bibinfo {year} {1995})}\BibitemShut
  {NoStop}%
\bibitem [{\citenamefont {Guo}\ \emph {et~al.}(2007)\citenamefont {Guo},
  \citenamefont {Yan}, \citenamefont {Wang}, \citenamefont {Tan}, \citenamefont
  {Zhang}, \citenamefont {Sui},\ and\ \citenamefont {Ma}}]{guo2007tensile}%
  \BibitemOpen
  \bibfield  {author} {\bibinfo {author} {\bibfnamefont {H.}~\bibnamefont
  {Guo}}, \bibinfo {author} {\bibfnamefont {P.}~\bibnamefont {Yan}}, \bibinfo
  {author} {\bibfnamefont {Y.}~\bibnamefont {Wang}}, \bibinfo {author}
  {\bibfnamefont {J.}~\bibnamefont {Tan}}, \bibinfo {author} {\bibfnamefont
  {Z.}~\bibnamefont {Zhang}}, \bibinfo {author} {\bibfnamefont
  {M.}~\bibnamefont {Sui}},\ and\ \bibinfo {author} {\bibfnamefont
  {E.}~\bibnamefont {Ma}},\ }\bibfield  {title} {\bibinfo {title} {Tensile
  ductility and necking of metallic glass},\ }\href
  {https://doi.org/https://doi.org/10.1038/nmat1984} {\bibfield  {journal}
  {\bibinfo  {journal} {Nat. Mater}\ }\textbf {\bibinfo {volume} {6}},\
  \bibinfo {pages} {735} (\bibinfo {year} {2007})}\BibitemShut {NoStop}%
\bibitem [{\citenamefont {Volkert}\ \emph {et~al.}(2008)\citenamefont
  {Volkert}, \citenamefont {Donohue},\ and\ \citenamefont
  {Spaepen}}]{volkert2008effect}%
  \BibitemOpen
  \bibfield  {author} {\bibinfo {author} {\bibfnamefont {C.}~\bibnamefont
  {Volkert}}, \bibinfo {author} {\bibfnamefont {A.}~\bibnamefont {Donohue}},\
  and\ \bibinfo {author} {\bibfnamefont {F.}~\bibnamefont {Spaepen}},\
  }\bibfield  {title} {\bibinfo {title} {Effect of sample size on deformation
  in amorphous metals},\ }\href
  {https://doi.org/https://doi.org/10.1063/1.2884584} {\bibfield  {journal}
  {\bibinfo  {journal} {J. Appl. Phys.}\ }\textbf {\bibinfo {volume} {103}},\
  \bibinfo {pages} {083539} (\bibinfo {year} {2008})}\BibitemShut {NoStop}%
\bibitem [{\citenamefont {Jang}\ and\ \citenamefont
  {Greer}(2010)}]{jang2010transition}%
  \BibitemOpen
  \bibfield  {author} {\bibinfo {author} {\bibfnamefont {D.}~\bibnamefont
  {Jang}}\ and\ \bibinfo {author} {\bibfnamefont {J.~R.}\ \bibnamefont
  {Greer}},\ }\bibfield  {title} {\bibinfo {title} {Transition from a
  strong-yet-brittle to a stronger-and-ductile state by size reduction of
  metallic glasses},\ }\href {https://doi.org/https://doi.org/10.1038/nmat2622}
  {\bibfield  {journal} {\bibinfo  {journal} {Nat. Mater}\ }\textbf {\bibinfo
  {volume} {9}},\ \bibinfo {pages} {215} (\bibinfo {year} {2010})}\BibitemShut
  {NoStop}%
\bibitem [{\citenamefont {Magagnosc}\ \emph {et~al.}(2013)\citenamefont
  {Magagnosc}, \citenamefont {Ehrbar}, \citenamefont {Kumar}, \citenamefont
  {He}, \citenamefont {Schroers},\ and\ \citenamefont
  {Gianola}}]{magagnosc2013tunable}%
  \BibitemOpen
  \bibfield  {author} {\bibinfo {author} {\bibfnamefont {D.}~\bibnamefont
  {Magagnosc}}, \bibinfo {author} {\bibfnamefont {R.}~\bibnamefont {Ehrbar}},
  \bibinfo {author} {\bibfnamefont {G.}~\bibnamefont {Kumar}}, \bibinfo
  {author} {\bibfnamefont {M.}~\bibnamefont {He}}, \bibinfo {author}
  {\bibfnamefont {J.}~\bibnamefont {Schroers}},\ and\ \bibinfo {author}
  {\bibfnamefont {D.}~\bibnamefont {Gianola}},\ }\bibfield  {title} {\bibinfo
  {title} {Tunable tensile ductility in metallic glasses},\ }\href
  {https://doi.org/https://doi.org/10.1038/srep01096} {\bibfield  {journal}
  {\bibinfo  {journal} {Scientific reports}\ }\textbf {\bibinfo {volume} {3}},\
  \bibinfo {pages} {1} (\bibinfo {year} {2013})}\BibitemShut {NoStop}%
\bibitem [{\citenamefont {Bonfanti}\ \emph {et~al.}(2018)\citenamefont
  {Bonfanti}, \citenamefont {Ferrero}, \citenamefont {Sellerio}, \citenamefont
  {Guerra},\ and\ \citenamefont {Zapperi}}]{bonfanti2018damage}%
  \BibitemOpen
  \bibfield  {author} {\bibinfo {author} {\bibfnamefont {S.}~\bibnamefont
  {Bonfanti}}, \bibinfo {author} {\bibfnamefont {E.~E.}\ \bibnamefont
  {Ferrero}}, \bibinfo {author} {\bibfnamefont {A.~L.}\ \bibnamefont
  {Sellerio}}, \bibinfo {author} {\bibfnamefont {R.}~\bibnamefont {Guerra}},\
  and\ \bibinfo {author} {\bibfnamefont {S.}~\bibnamefont {Zapperi}},\
  }\bibfield  {title} {\bibinfo {title} {Damage accumulation in silica glass
  nanofibers},\ }\href
  {https://doi.org/https://doi.org/10.1021/acs.nanolett.8b00469} {\bibfield
  {journal} {\bibinfo  {journal} {Nano Letters}\ }\textbf {\bibinfo {volume}
  {18}},\ \bibinfo {pages} {4100} (\bibinfo {year} {2018})}\BibitemShut
  {NoStop}%
\bibitem [{\citenamefont {Shi}(2019)}]{shi2019size}%
  \BibitemOpen
  \bibfield  {author} {\bibinfo {author} {\bibfnamefont {Y.}~\bibnamefont
  {Shi}},\ }\bibfield  {title} {\bibinfo {title} {Size-dependent mechanical
  responses of metallic glasses},\ }\href
  {https://doi.org/https://doi.org/10.1080/09506608.2018.1476079} {\bibfield
  {journal} {\bibinfo  {journal} {Int. Mater. Rev.}\ }\textbf {\bibinfo
  {volume} {64}},\ \bibinfo {pages} {163} (\bibinfo {year} {2019})}\BibitemShut
  {NoStop}%
\bibitem [{\citenamefont {Falk}\ and\ \citenamefont
  {Langer}(1998)}]{falk1998dynamics}%
  \BibitemOpen
  \bibfield  {author} {\bibinfo {author} {\bibfnamefont {M.~L.}\ \bibnamefont
  {Falk}}\ and\ \bibinfo {author} {\bibfnamefont {J.~S.}\ \bibnamefont
  {Langer}},\ }\bibfield  {title} {\bibinfo {title} {Dynamics of viscoplastic
  deformation in amorphous solids},\ }\href
  {https://doi.org/https://doi.org/10.1103/PhysRevE.57.7192} {\bibfield
  {journal} {\bibinfo  {journal} {Phys. Rev. E}\ }\textbf {\bibinfo {volume}
  {57}},\ \bibinfo {pages} {7192} (\bibinfo {year} {1998})}\BibitemShut
  {NoStop}%
\bibitem [{\citenamefont {Menzl}\ \emph {et~al.}(2016)\citenamefont {Menzl},
  \citenamefont {Gonzalez}, \citenamefont {Geiger}, \citenamefont {Caupin},
  \citenamefont {Abascal}, \citenamefont {Valeriani},\ and\ \citenamefont
  {Dellago}}]{menzl2016molecular}%
  \BibitemOpen
  \bibfield  {author} {\bibinfo {author} {\bibfnamefont {G.}~\bibnamefont
  {Menzl}}, \bibinfo {author} {\bibfnamefont {M.~A.}\ \bibnamefont {Gonzalez}},
  \bibinfo {author} {\bibfnamefont {P.}~\bibnamefont {Geiger}}, \bibinfo
  {author} {\bibfnamefont {F.}~\bibnamefont {Caupin}}, \bibinfo {author}
  {\bibfnamefont {J.~L.}\ \bibnamefont {Abascal}}, \bibinfo {author}
  {\bibfnamefont {C.}~\bibnamefont {Valeriani}},\ and\ \bibinfo {author}
  {\bibfnamefont {C.}~\bibnamefont {Dellago}},\ }\bibfield  {title} {\bibinfo
  {title} {Molecular mechanism for cavitation in water under tension},\ }\href
  {https://doi.org/https://doi.org/10.1073/pnas.1608421113} {\bibfield
  {journal} {\bibinfo  {journal} {Proc. Natl. Acad. Sci. U.S.A.}\ }\textbf
  {\bibinfo {volume} {113}},\ \bibinfo {pages} {13582} (\bibinfo {year}
  {2016})}\BibitemShut {NoStop}%
\bibitem [{\citenamefont {Galimzyanov}\ and\ \citenamefont
  {Mokshin}(2022)}]{galimzyanov2022cavity}%
  \BibitemOpen
  \bibfield  {author} {\bibinfo {author} {\bibfnamefont {B.}~\bibnamefont
  {Galimzyanov}}\ and\ \bibinfo {author} {\bibfnamefont {A.}~\bibnamefont
  {Mokshin}},\ }\bibfield  {title} {\bibinfo {title} {Cavity nucleation in
  single-component homogeneous amorphous solids under negative pressure},\
  }\href {https://doi.org/10.1088/1361-648X/ac8462} {\bibfield  {journal}
  {\bibinfo  {journal} {J. Condens. Matter Phys.}\ }\textbf {\bibinfo {volume}
  {34}},\ \bibinfo {pages} {414001} (\bibinfo {year} {2022})}\BibitemShut
  {NoStop}%
\bibitem [{\citenamefont {Tang}\ \emph {et~al.}(2020)\citenamefont {Tang},
  \citenamefont {Nguyen}, \citenamefont {Yao},\ and\ \citenamefont
  {Wilkerson}}]{tang2020cavitation}%
  \BibitemOpen
  \bibfield  {author} {\bibinfo {author} {\bibfnamefont {X.}~\bibnamefont
  {Tang}}, \bibinfo {author} {\bibfnamefont {T.}~\bibnamefont {Nguyen}},
  \bibinfo {author} {\bibfnamefont {X.}~\bibnamefont {Yao}},\ and\ \bibinfo
  {author} {\bibfnamefont {J.~W.}\ \bibnamefont {Wilkerson}},\ }\bibfield
  {title} {\bibinfo {title} {A cavitation and dynamic void growth model for a
  general class of strain-softening amorphous materials},\ }\href
  {https://doi.org/https://doi.org/10.1016/j.jmps.2020.104023} {\bibfield
  {journal} {\bibinfo  {journal} {J. Mech. Phys. Solids.}\ }\textbf {\bibinfo
  {volume} {141}},\ \bibinfo {pages} {104023} (\bibinfo {year}
  {2020})}\BibitemShut {NoStop}%
\bibitem [{\citenamefont {Lewandowski}\ and\ \citenamefont
  {Greer}(2006)}]{lewandowski2006temperature}%
  \BibitemOpen
  \bibfield  {author} {\bibinfo {author} {\bibfnamefont {J.}~\bibnamefont
  {Lewandowski}}\ and\ \bibinfo {author} {\bibfnamefont {A.}~\bibnamefont
  {Greer}},\ }\bibfield  {title} {\bibinfo {title} {Temperature rise at shear
  bands in metallic glasses},\ }\href
  {https://doi.org/https://doi.org/10.1038/nmat1536} {\bibfield  {journal}
  {\bibinfo  {journal} {Nat. Mater}\ }\textbf {\bibinfo {volume} {5}},\
  \bibinfo {pages} {15} (\bibinfo {year} {2006})}\BibitemShut {NoStop}%
\bibitem [{\citenamefont {Mota}\ \emph {et~al.}(2021)\citenamefont {Mota},
  \citenamefont {Lund}, \citenamefont {Sohn}, \citenamefont {Browne},
  \citenamefont {Hofmann}, \citenamefont {Curtarolo}, \citenamefont {van~de
  Walle},\ and\ \citenamefont {Schroers}}]{mota2021enhancing}%
  \BibitemOpen
  \bibfield  {author} {\bibinfo {author} {\bibfnamefont {R.~M.~O.}\
  \bibnamefont {Mota}}, \bibinfo {author} {\bibfnamefont {E.~T.}\ \bibnamefont
  {Lund}}, \bibinfo {author} {\bibfnamefont {S.}~\bibnamefont {Sohn}}, \bibinfo
  {author} {\bibfnamefont {D.~J.}\ \bibnamefont {Browne}}, \bibinfo {author}
  {\bibfnamefont {D.~C.}\ \bibnamefont {Hofmann}}, \bibinfo {author}
  {\bibfnamefont {S.}~\bibnamefont {Curtarolo}}, \bibinfo {author}
  {\bibfnamefont {A.}~\bibnamefont {van~de Walle}},\ and\ \bibinfo {author}
  {\bibfnamefont {J.}~\bibnamefont {Schroers}},\ }\bibfield  {title} {\bibinfo
  {title} {Enhancing ductility in bulk metallic glasses by straining during
  cooling},\ }\href
  {https://doi.org/https://doi.org/10.1038/s43246-021-00127-0} {\bibfield
  {journal} {\bibinfo  {journal} {Commun. Mater}\ }\textbf {\bibinfo {volume}
  {2}},\ \bibinfo {pages} {1} (\bibinfo {year} {2021})}\BibitemShut {NoStop}%
\bibitem [{\citenamefont {Liu}\ \emph {et~al.}(2017)\citenamefont {Liu},
  \citenamefont {Roddatis}, \citenamefont {Kenesei},\ and\ \citenamefont
  {Maa{\ss}}}]{liu2017shear}%
  \BibitemOpen
  \bibfield  {author} {\bibinfo {author} {\bibfnamefont {C.}~\bibnamefont
  {Liu}}, \bibinfo {author} {\bibfnamefont {V.}~\bibnamefont {Roddatis}},
  \bibinfo {author} {\bibfnamefont {P.}~\bibnamefont {Kenesei}},\ and\ \bibinfo
  {author} {\bibfnamefont {R.}~\bibnamefont {Maa{\ss}}},\ }\bibfield  {title}
  {\bibinfo {title} {Shear-band thickness and shear-band cavities in a zr-based
  metallic glass},\ }\href
  {https://doi.org/https://doi.org/10.1016/j.actamat.2017.08.032} {\bibfield
  {journal} {\bibinfo  {journal} {Acta Mater.}\ }\textbf {\bibinfo {volume}
  {140}},\ \bibinfo {pages} {206} (\bibinfo {year} {2017})}\BibitemShut
  {NoStop}%
\bibitem [{\citenamefont {Lu}\ \emph {et~al.}(2018)\citenamefont {Lu},
  \citenamefont {Jiang}, \citenamefont {Lu}, \citenamefont {Qin}, \citenamefont
  {Huang},\ and\ \citenamefont {Shen}}]{lu2018dilatancy}%
  \BibitemOpen
  \bibfield  {author} {\bibinfo {author} {\bibfnamefont {Y.}~\bibnamefont
  {Lu}}, \bibinfo {author} {\bibfnamefont {M.}~\bibnamefont {Jiang}}, \bibinfo
  {author} {\bibfnamefont {X.}~\bibnamefont {Lu}}, \bibinfo {author}
  {\bibfnamefont {Z.}~\bibnamefont {Qin}}, \bibinfo {author} {\bibfnamefont
  {Y.}~\bibnamefont {Huang}},\ and\ \bibinfo {author} {\bibfnamefont
  {J.}~\bibnamefont {Shen}},\ }\bibfield  {title} {\bibinfo {title} {Dilatancy
  of shear transformations in a colloidal glass},\ }\href
  {https://doi.org/https://doi.org/10.1103/PhysRevApplied.9.014023} {\bibfield
  {journal} {\bibinfo  {journal} {Phys. Rev. Appl.}\ }\textbf {\bibinfo
  {volume} {9}},\ \bibinfo {pages} {014023} (\bibinfo {year}
  {2018})}\BibitemShut {NoStop}%
\bibitem [{\citenamefont {Wang}\ \emph {et~al.}(2022)\citenamefont {Wang},
  \citenamefont {Lu}, \citenamefont {Lu}, \citenamefont {Huo}, \citenamefont
  {Wang}, \citenamefont {Wang}, \citenamefont {Dai},\ and\ \citenamefont
  {Jiang}}]{wang2022elastic}%
  \BibitemOpen
  \bibfield  {author} {\bibinfo {author} {\bibfnamefont {X.}~\bibnamefont
  {Wang}}, \bibinfo {author} {\bibfnamefont {Y.}~\bibnamefont {Lu}}, \bibinfo
  {author} {\bibfnamefont {X.}~\bibnamefont {Lu}}, \bibinfo {author}
  {\bibfnamefont {J.}~\bibnamefont {Huo}}, \bibinfo {author} {\bibfnamefont
  {Y.}~\bibnamefont {Wang}}, \bibinfo {author} {\bibfnamefont {W.}~\bibnamefont
  {Wang}}, \bibinfo {author} {\bibfnamefont {L.}~\bibnamefont {Dai}},\ and\
  \bibinfo {author} {\bibfnamefont {M.}~\bibnamefont {Jiang}},\ }\bibfield
  {title} {\bibinfo {title} {Elastic criterion for shear-banding instability in
  amorphous solids},\ }\href
  {https://doi.org/https://doi.org/10.1103/PhysRevE.105.045003} {\bibfield
  {journal} {\bibinfo  {journal} {Phys. Rev. E}\ }\textbf {\bibinfo {volume}
  {105}},\ \bibinfo {pages} {045003} (\bibinfo {year} {2022})}\BibitemShut
  {NoStop}%
\bibitem [{\citenamefont {Klaum{\"u}nzer}\ \emph {et~al.}(2011)\citenamefont
  {Klaum{\"u}nzer}, \citenamefont {Lazarev}, \citenamefont {Maa{\ss}},
  \citenamefont {Dalla~Torre}, \citenamefont {Vinogradov},\ and\ \citenamefont
  {L{\"o}ffler}}]{klaumunzer2011probing}%
  \BibitemOpen
  \bibfield  {author} {\bibinfo {author} {\bibfnamefont {D.}~\bibnamefont
  {Klaum{\"u}nzer}}, \bibinfo {author} {\bibfnamefont {A.}~\bibnamefont
  {Lazarev}}, \bibinfo {author} {\bibfnamefont {R.}~\bibnamefont {Maa{\ss}}},
  \bibinfo {author} {\bibfnamefont {F.}~\bibnamefont {Dalla~Torre}}, \bibinfo
  {author} {\bibfnamefont {A.}~\bibnamefont {Vinogradov}},\ and\ \bibinfo
  {author} {\bibfnamefont {J.~F.}\ \bibnamefont {L{\"o}ffler}},\ }\bibfield
  {title} {\bibinfo {title} {Probing shear-band initiation in metallic
  glasses},\ }\href
  {https://doi.org/https://doi.org/10.1103/PhysRevLett.107.185502} {\bibfield
  {journal} {\bibinfo  {journal} {Phys. Rev. Lett.}\ }\textbf {\bibinfo
  {volume} {107}},\ \bibinfo {pages} {185502} (\bibinfo {year}
  {2011})}\BibitemShut {NoStop}%
\bibitem [{\citenamefont {Murali}\ \emph {et~al.}(2011)\citenamefont {Murali},
  \citenamefont {Guo}, \citenamefont {Zhang}, \citenamefont {Narasimhan},
  \citenamefont {Li},\ and\ \citenamefont {Gao}}]{murali2011atomic}%
  \BibitemOpen
  \bibfield  {author} {\bibinfo {author} {\bibfnamefont {P.}~\bibnamefont
  {Murali}}, \bibinfo {author} {\bibfnamefont {T.}~\bibnamefont {Guo}},
  \bibinfo {author} {\bibfnamefont {Y.}~\bibnamefont {Zhang}}, \bibinfo
  {author} {\bibfnamefont {R.}~\bibnamefont {Narasimhan}}, \bibinfo {author}
  {\bibfnamefont {Y.}~\bibnamefont {Li}},\ and\ \bibinfo {author}
  {\bibfnamefont {H.}~\bibnamefont {Gao}},\ }\bibfield  {title} {\bibinfo
  {title} {Atomic scale fluctuations govern brittle fracture and cavitation
  behavior in metallic glasses},\ }\href
  {https://doi.org/https://doi.org/10.1103/PhysRevLett.107.215501} {\bibfield
  {journal} {\bibinfo  {journal} {Phys. Rev. Lett.}\ }\textbf {\bibinfo
  {volume} {107}},\ \bibinfo {pages} {215501} (\bibinfo {year}
  {2011})}\BibitemShut {NoStop}%
\bibitem [{\citenamefont {Rycroft}\ and\ \citenamefont
  {Bouchbinder}(2012)}]{rycroft2012fracture}%
  \BibitemOpen
  \bibfield  {author} {\bibinfo {author} {\bibfnamefont {C.~H.}\ \bibnamefont
  {Rycroft}}\ and\ \bibinfo {author} {\bibfnamefont {E.}~\bibnamefont
  {Bouchbinder}},\ }\bibfield  {title} {\bibinfo {title} {Fracture toughness of
  metallic glasses: Annealing-induced embrittlement},\ }\href
  {https://doi.org/10.1103/PhysRevLett.109.194301} {\bibfield  {journal}
  {\bibinfo  {journal} {Phys. Rev. Lett.}\ }\textbf {\bibinfo {volume} {109}},\
  \bibinfo {pages} {194301} (\bibinfo {year} {2012})}\BibitemShut {NoStop}%
\bibitem [{\citenamefont {Guan}\ \emph {et~al.}(2013)\citenamefont {Guan},
  \citenamefont {Lu}, \citenamefont {Spector}, \citenamefont {Valavala},\ and\
  \citenamefont {Falk}}]{guan2013cavitation}%
  \BibitemOpen
  \bibfield  {author} {\bibinfo {author} {\bibfnamefont {P.}~\bibnamefont
  {Guan}}, \bibinfo {author} {\bibfnamefont {S.}~\bibnamefont {Lu}}, \bibinfo
  {author} {\bibfnamefont {M.~J.}\ \bibnamefont {Spector}}, \bibinfo {author}
  {\bibfnamefont {P.~K.}\ \bibnamefont {Valavala}},\ and\ \bibinfo {author}
  {\bibfnamefont {M.~L.}\ \bibnamefont {Falk}},\ }\bibfield  {title} {\bibinfo
  {title} {Cavitation in amorphous solids},\ }\href
  {https://doi.org/https://doi.org/10.1103/PhysRevLett.110.185502} {\bibfield
  {journal} {\bibinfo  {journal} {Phys. Rev. Lett.}\ }\textbf {\bibinfo
  {volume} {110}},\ \bibinfo {pages} {185502} (\bibinfo {year}
  {2013})}\BibitemShut {NoStop}%
\bibitem [{\citenamefont {Vasoya}\ \emph {et~al.}(2016)\citenamefont {Vasoya},
  \citenamefont {Rycroft},\ and\ \citenamefont
  {Bouchbinder}}]{vasoya2016notch}%
  \BibitemOpen
  \bibfield  {author} {\bibinfo {author} {\bibfnamefont {M.}~\bibnamefont
  {Vasoya}}, \bibinfo {author} {\bibfnamefont {C.~H.}\ \bibnamefont
  {Rycroft}},\ and\ \bibinfo {author} {\bibfnamefont {E.}~\bibnamefont
  {Bouchbinder}},\ }\bibfield  {title} {\bibinfo {title} {Notch fracture
  toughness of glasses: Dependence on rate, age, and geometry},\ }\href
  {https://doi.org/10.1103/PhysRevApplied.6.024008} {\bibfield  {journal}
  {\bibinfo  {journal} {Phys. Rev. Applied}\ }\textbf {\bibinfo {volume} {6}},\
  \bibinfo {pages} {024008} (\bibinfo {year} {2016})}\BibitemShut {NoStop}%
\bibitem [{\citenamefont {Falk}(1999)}]{falk1999molecular}%
  \BibitemOpen
  \bibfield  {author} {\bibinfo {author} {\bibfnamefont {M.}~\bibnamefont
  {Falk}},\ }\bibfield  {title} {\bibinfo {title} {Molecular-dynamics study of
  ductile and brittle fracture in model noncrystalline solids},\ }\href
  {https://doi.org/https://doi.org/10.1103/PhysRevB.60.7062} {\bibfield
  {journal} {\bibinfo  {journal} {Phys. Rev. B}\ }\textbf {\bibinfo {volume}
  {60}},\ \bibinfo {pages} {7062} (\bibinfo {year} {1999})}\BibitemShut
  {NoStop}%
\bibitem [{\citenamefont {Murali}\ \emph {et~al.}(2013)\citenamefont {Murali},
  \citenamefont {Narasimhan}, \citenamefont {Guo}, \citenamefont {Zhang},\ and\
  \citenamefont {Gao}}]{murali2013shear}%
  \BibitemOpen
  \bibfield  {author} {\bibinfo {author} {\bibfnamefont {P.}~\bibnamefont
  {Murali}}, \bibinfo {author} {\bibfnamefont {R.}~\bibnamefont {Narasimhan}},
  \bibinfo {author} {\bibfnamefont {T.}~\bibnamefont {Guo}}, \bibinfo {author}
  {\bibfnamefont {Y.}~\bibnamefont {Zhang}},\ and\ \bibinfo {author}
  {\bibfnamefont {H.}~\bibnamefont {Gao}},\ }\bibfield  {title} {\bibinfo
  {title} {Shear bands mediate cavitation in brittle metallic glasses},\ }\href
  {https://doi.org/https://doi.org/10.1016/j.scriptamat.2012.11.038} {\bibfield
   {journal} {\bibinfo  {journal} {Scr. Mater.}\ }\textbf {\bibinfo {volume}
  {68}},\ \bibinfo {pages} {567} (\bibinfo {year} {2013})}\BibitemShut
  {NoStop}%
\bibitem [{\citenamefont {Luo}\ and\ \citenamefont
  {Shi}(2015)}]{luo2015tensile}%
  \BibitemOpen
  \bibfield  {author} {\bibinfo {author} {\bibfnamefont {J.}~\bibnamefont
  {Luo}}\ and\ \bibinfo {author} {\bibfnamefont {Y.}~\bibnamefont {Shi}},\
  }\bibfield  {title} {\bibinfo {title} {Tensile fracture of metallic glasses
  via shear band cavitation},\ }\href
  {https://doi.org/https://doi.org/10.1016/j.actamat.2014.09.008} {\bibfield
  {journal} {\bibinfo  {journal} {Acta Mater.}\ }\textbf {\bibinfo {volume}
  {82}},\ \bibinfo {pages} {483} (\bibinfo {year} {2015})}\BibitemShut
  {NoStop}%
\bibitem [{\citenamefont {Shao}\ \emph {et~al.}(2014)\citenamefont {Shao},
  \citenamefont {Yang}, \citenamefont {Yao},\ and\ \citenamefont
  {Liu}}]{shao2014direct}%
  \BibitemOpen
  \bibfield  {author} {\bibinfo {author} {\bibfnamefont {Y.}~\bibnamefont
  {Shao}}, \bibinfo {author} {\bibfnamefont {G.-N.}\ \bibnamefont {Yang}},
  \bibinfo {author} {\bibfnamefont {K.-F.}\ \bibnamefont {Yao}},\ and\ \bibinfo
  {author} {\bibfnamefont {X.}~\bibnamefont {Liu}},\ }\bibfield  {title}
  {\bibinfo {title} {Direct experimental evidence of nano-voids formation and
  coalescence within shear bands},\ }\href
  {https://doi.org/https://doi.org/10.1063/1.4901281} {\bibfield  {journal}
  {\bibinfo  {journal} {Appl. Phys. Lett.}\ }\textbf {\bibinfo {volume}
  {105}},\ \bibinfo {pages} {181909} (\bibinfo {year} {2014})}\BibitemShut
  {NoStop}%
\bibitem [{\citenamefont {Wright}\ \emph {et~al.}(2003)\citenamefont {Wright},
  \citenamefont {Hufnagel},\ and\ \citenamefont {Nix}}]{wright2003free}%
  \BibitemOpen
  \bibfield  {author} {\bibinfo {author} {\bibfnamefont {W.~J.}\ \bibnamefont
  {Wright}}, \bibinfo {author} {\bibfnamefont {T.}~\bibnamefont {Hufnagel}},\
  and\ \bibinfo {author} {\bibfnamefont {W.}~\bibnamefont {Nix}},\ }\bibfield
  {title} {\bibinfo {title} {Free volume coalescence and void formation in
  shear bands in metallic glass},\ }\href
  {https://doi.org/https://doi.org/10.1063/1.1531212} {\bibfield  {journal}
  {\bibinfo  {journal} {J. Appl. Phys.}\ }\textbf {\bibinfo {volume} {93}},\
  \bibinfo {pages} {1432} (\bibinfo {year} {2003})}\BibitemShut {NoStop}%
\bibitem [{\citenamefont {Shi}(2016)}]{shi2016creating}%
  \BibitemOpen
  \bibfield  {author} {\bibinfo {author} {\bibfnamefont {Y.}~\bibnamefont
  {Shi}},\ }\bibfield  {title} {\bibinfo {title} {Creating atomic models of
  brittle glasses for in silico mechanical tests},\ }\href
  {https://doi.org/https://doi.org/10.1111/ijag.12253} {\bibfield  {journal}
  {\bibinfo  {journal} {Int. J. Appl. Glass Sci.}\ }\textbf {\bibinfo {volume}
  {7}},\ \bibinfo {pages} {464} (\bibinfo {year} {2016})}\BibitemShut {NoStop}%
\bibitem [{\citenamefont {Zhao}\ \emph {et~al.}(2023)\citenamefont {Zhao},
  \citenamefont {Wang},\ and\ \citenamefont {Cao}}]{zhao2023fracture}%
  \BibitemOpen
  \bibfield  {author} {\bibinfo {author} {\bibfnamefont {K.}~\bibnamefont
  {Zhao}}, \bibinfo {author} {\bibfnamefont {Y.-J.}\ \bibnamefont {Wang}},\
  and\ \bibinfo {author} {\bibfnamefont {P.}~\bibnamefont {Cao}},\ }\bibfield
  {title} {\bibinfo {title} {Fracture universality in amorphous nanowires},\
  }\href {https://doi.org/https://doi.org/10.1016/j.jmps.2023.105210}
  {\bibfield  {journal} {\bibinfo  {journal} {J. Mech. Phys. Solids.}\ }\textbf
  {\bibinfo {volume} {173}},\ \bibinfo {pages} {105210} (\bibinfo {year}
  {2023})}\BibitemShut {NoStop}%
\bibitem [{\citenamefont {He}\ \emph {et~al.}(2019)\citenamefont {He},
  \citenamefont {Yi},\ and\ \citenamefont {Falk}}]{he2019critical}%
  \BibitemOpen
  \bibfield  {author} {\bibinfo {author} {\bibfnamefont {Y.}~\bibnamefont
  {He}}, \bibinfo {author} {\bibfnamefont {P.}~\bibnamefont {Yi}},\ and\
  \bibinfo {author} {\bibfnamefont {M.~L.}\ \bibnamefont {Falk}},\ }\bibfield
  {title} {\bibinfo {title} {Critical analysis of an fep empirical potential
  employed to study the fracture of metallic glasses},\ }\href
  {https://doi.org/https://doi.org/10.1103/PhysRevLett.122.035501} {\bibfield
  {journal} {\bibinfo  {journal} {Phys. Rev. Lett.}\ }\textbf {\bibinfo
  {volume} {122}},\ \bibinfo {pages} {035501} (\bibinfo {year}
  {2019})}\BibitemShut {NoStop}%
\bibitem [{\citenamefont {Tang}\ \emph {et~al.}(2022)\citenamefont {Tang},
  \citenamefont {Shen}, \citenamefont {Zhang}, \citenamefont {Li},\ and\
  \citenamefont {Wang}}]{tang2022crack}%
  \BibitemOpen
  \bibfield  {author} {\bibinfo {author} {\bibfnamefont {X.}~\bibnamefont
  {Tang}}, \bibinfo {author} {\bibfnamefont {L.}~\bibnamefont {Shen}}, \bibinfo
  {author} {\bibfnamefont {H.}~\bibnamefont {Zhang}}, \bibinfo {author}
  {\bibfnamefont {W.}~\bibnamefont {Li}},\ and\ \bibinfo {author}
  {\bibfnamefont {W.}~\bibnamefont {Wang}},\ }\bibfield  {title} {\bibinfo
  {title} {Crack tip cavitation in metallic glasses},\ }\href
  {https://doi.org/https://doi.org/10.1016/j.jnoncrysol.2022.121762} {\bibfield
   {journal} {\bibinfo  {journal} {J. Non-Cryst. Solids}\ }\textbf {\bibinfo
  {volume} {592}},\ \bibinfo {pages} {121762} (\bibinfo {year}
  {2022})}\BibitemShut {NoStop}%
\bibitem [{\citenamefont {Lerner}(2019)}]{lerner2019mechanical}%
  \BibitemOpen
  \bibfield  {author} {\bibinfo {author} {\bibfnamefont {E.}~\bibnamefont
  {Lerner}},\ }\bibfield  {title} {\bibinfo {title} {Mechanical properties of
  simple computer glasses},\ }\href
  {https://doi.org/https://doi.org/10.1016/j.jnoncrysol.2019.119570} {\bibfield
   {journal} {\bibinfo  {journal} {J. Non-Cryst. Solids}\ }\textbf {\bibinfo
  {volume} {522}},\ \bibinfo {pages} {119570} (\bibinfo {year}
  {2019})}\BibitemShut {NoStop}%
\bibitem [{\citenamefont {P{\u{a}}duraru}\ \emph {et~al.}(2010)\citenamefont
  {P{\u{a}}duraru}, \citenamefont {Andersen}, \citenamefont {Thyssen},
  \citenamefont {Bailey}, \citenamefont {Jacobsen},\ and\ \citenamefont
  {Schi{\o}tz}}]{puaduraru2010computer}%
  \BibitemOpen
  \bibfield  {author} {\bibinfo {author} {\bibfnamefont {A.}~\bibnamefont
  {P{\u{a}}duraru}}, \bibinfo {author} {\bibfnamefont {U.~G.}\ \bibnamefont
  {Andersen}}, \bibinfo {author} {\bibfnamefont {A.}~\bibnamefont {Thyssen}},
  \bibinfo {author} {\bibfnamefont {N.}~\bibnamefont {Bailey}}, \bibinfo
  {author} {\bibfnamefont {K.~W.}\ \bibnamefont {Jacobsen}},\ and\ \bibinfo
  {author} {\bibfnamefont {J.}~\bibnamefont {Schi{\o}tz}},\ }\bibfield  {title}
  {\bibinfo {title} {Computer simulations of nanoindentation in mg--cu and
  cu--zr metallic glasses},\ }\href
  {https://doi.org/10.1088/0965-0393/18/5/055006} {\bibfield  {journal}
  {\bibinfo  {journal} {Model. Simul. Mater. Sci. Eng.}\ }\textbf {\bibinfo
  {volume} {18}},\ \bibinfo {pages} {055006} (\bibinfo {year}
  {2010})}\BibitemShut {NoStop}%
\end{thebibliography}

%

\onecolumngrid
\newpage
\begin{center}
   \textbf{\large Supplementary material: ``Bridging necking and shear-banding mediated tensile failure in glasses''}
\end{center}

The goal of this document is to provide additional technical details regarding the results presented in the main text and some supporting results.

\section{Sample Preparation and uniaxial tension tests}

The metallic glass Zr$_{\mbox{\tiny 44}}$Ti$_{\mbox{\tiny 11}}$Cu$_{\mbox{\tiny 10}}$Ni$_{\mbox{\tiny 10}}$Be$_{\mbox{\tiny 25}}$ rods of 1.8 mm in diameter were prepared by copper mold casting. High purity ingots (the purity was higher than 99.99\% for all constituent elements) were arc-melted under argon atmosphere to alloy the constituents. Subsequently, the alloy was reheated to a temperature of $\sim\!1000$ $^\circ$C and  was forced through a quartz nozzle under an argon gas pressure into a copper mold. Thermal analysis on some of the rods revealed a typical thermogram characteristic of a fully amorphous structure, which was cooled at a rate of $\sim\!500$ K/s~\cite{ketkaew2018mechanical}.

\begin{figure}[h!]
\includegraphics[width = 0.9\textwidth]{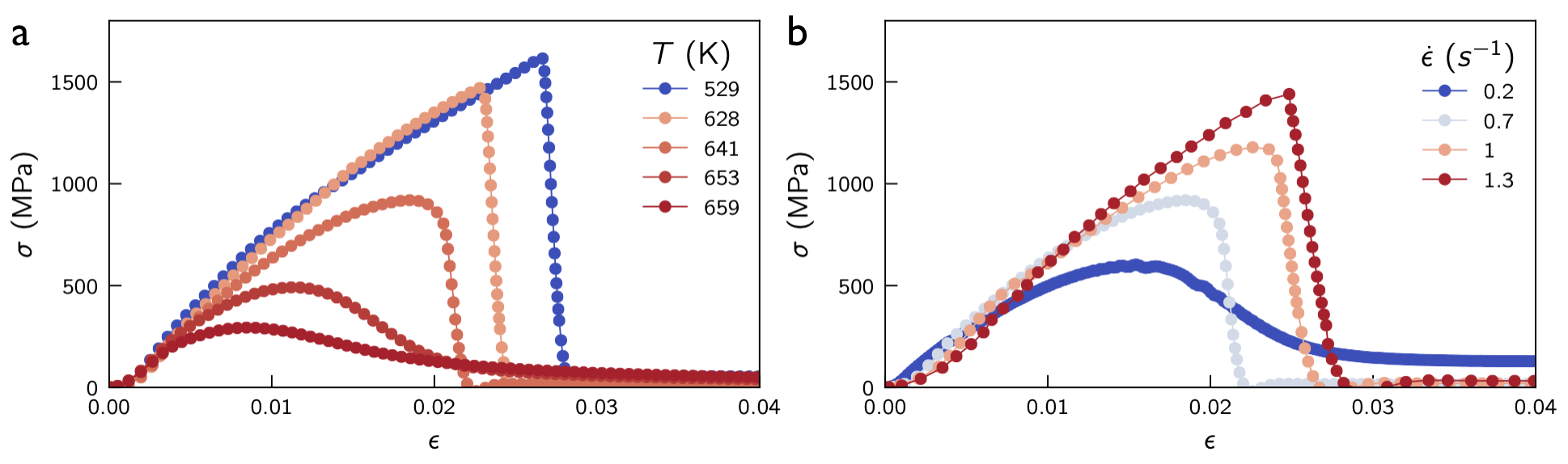}
\caption{\footnotesize (a) Stress-strain curves $\sigma(\epsilon)$ for various temperatures $T$ (see legend and note that $T_{\rm g}\!=\!625$ K) at a fixed engineering strain-rate $\dot{\epsilon}\!=\!0.7$ $s^{-1}$. (b) The same as in (a), but for various $\dot\epsilon$ (see legend) and a fixed temperature $T\!=\!641$ K. The ultimate tensile strength $\sigma_{\mbox{\tiny UTS}}(\dot\epsilon,T)$ is defined as the maximum of each $\sigma(\epsilon)$ curve.}
\label{fig:sm_exp}
\end{figure}

Samples were mounted onto a screw driven universal testing machine (Instron 5569), with a 30 kN load cell affixed with tension clamps, and a 5 mm section was heated to the target temperature via RF coil (Ameritherm EasyHeat), with a 40 s pre-heat. The temperature was set by RF coil power, where a power-temperature calibration was determined for this specific sample set using an N-type thermocouple and noncontact pyrometer. After pre-heating, loading rate controlled tension tests were initiated ($1.0\!-\!6.5$ mm/s, corresponding to the intended engineering strain-rates $\dot\epsilon$), halting shortly after material failure. The applied force and displacement measurements were used to generate the engineering stress-strain curves $\sigma(\epsilon)$ shown in Fig.~\ref{fig:sm_exp}a-b.

\section{Scanning electron microscope imaging and analysis}

The fracture surface of samples that featured a finite cross-sectional area at failure were imaged at 5.0 kV and 60x magnification, using scanning electron microscopy (Hitachi SU7000). To quantify the surface area covered by dimples vs.~veins (fractographic traces of dilation-driven cavities vs.~shear-driven plasticity, respectively), images were analyzed using the software ImageJ. Dimples and veins regions were detected by visual inspection, with dimples exhibiting a highly 3-dimensional structure and veins laying comparatively flat on the fractured surface. The boundaries between these surface patterns appear to be sharp, as shown in Fig.~1b in the main text. To measure surface area, both the total exposed fracture surface and fracture surface exhibiting dimples were quantified using oval and freehand selections, with the measurement scale set by the scale bar in the SEM image. The dimples area fraction (cf.~Fig.~1c in the main text) was determined as the area exhibiting dimples divided by the total exposed fracture surface. The maximum dimple size (cf.~Fig.~1d in the main text) was measured as the diameter of the largest dimple on a given fracture surface.

Samples that featured vanishingly small cross sections at failure (necking-mediated failure) were polished along rods' long axes. To this end, samples were mounted in a resin puck for stability and ease of use, then polished to a mirror finish using sandpaper and suspended colloidal solutions down to 3 $\mu$m grit. The samples were then imaged at 5.0 kV and 60x magnification, using scanning electron microscopy (Hitachi SU7000), as above.

\section{Large-scale molecular dynamics simulations}

A natural choice for simulating bulk metallic glasses might be to employ an embedded-atom method (EAM) potential. However, EAM force fields for BMGs are still not sufficiently robust to model all physical properties of realistic laboratory glasses. Most of the currently available EAM potentials for BMGs are notoriously ductile, featuring larger Poisson's ratio compared with laboratory glasses~\cite{shi2016creating}. As the quenching rates accessible through cutting-edge large-scale MD simulations are significantly larger than their laboratory counterparts, these models typically reveal extensive shear flow and necking (see, e.g.,~\cite{zhao2023fracture}, for Zr-based alloys). Moreover, while there exists an EAM potential for an FeP metallic glass that exhibits brittle behavior at high tensile stresses, it was recently demonstrated in~\cite{he2019critical} that spurious phase separation occurs during annealing, causing unphysical embrittlement. As an alternative, part of the community have developed modified LJ or Dzugurov-based potentials~\cite{luo2015tensile,shi2016creating,richard2021brittle,tang2022crack}, which enable the creation of good glass formers (no demixing, neither crystallization), while displaying both ductile and brittle behaviors. Those models show identical features of what has been seen using EAM potentials~\cite{murali2011atomic}, e.g.~void formation near a crack tip during embrittlement~\cite{richard2021brittle,tang2022crack}. We follow this route in this work.

Our computer glasses are composed of a 50:50 binary mixture. Particles of equal mass m interact with a modified Lennard-Jones type potential of the form
\begin{equation}
    \varphi(r_{ij},\lambda_{ij}) \!=\!
\left\{
\begin{array}{cc}
\!\!4\varepsilon \bigg[ \big(\frac{\lambda_{ij}}{r_{ij}}\big)^{12} - \big(\frac{\lambda_{ij}}{r_{ij}}\big) ^{6} \bigg],
     &  \frac{r_{ij}}{\lambda_{ij}} <  x_{\mbox{\tiny min}}  \\
\varepsilon \bigg[a\big(\frac{\lambda_{ij}}{r_{ij}}\big)^{12} -b\big(\frac{\lambda_{ij}}{r_{ij}}\big)^{6} + \sum\limits_{\ell=0} ^{3}  c_{\mbox{\tiny $2\ell$}} \big(\frac{r_{ij}}{\lambda_{ij}}\big)^{2\ell} \bigg] , & x_{\mbox{\tiny min}}\!\le\! \frac{r_{ij}}{\lambda_{ij}}< x_c\\
0\,,  & x_c \le \frac{r_{ij}}{\lambda_{ij}}
\end{array}
\right. ,
 \label{eq:potential}
\end{equation}
where $\varepsilon$ is a microscopic energy scale (not to be confused with the engineering strain $\epsilon$), $x_{\mbox{\tiny min}},x_c$ are the (dimensionless) locations of the minimum of the Lennard-Jones potential and modified cutoff, respectively, and the $\lambda_{ij}$'s are the length parameters, further described below. We express the dimensionless cutoff $x_c$ in terms of $x_{\mbox{\tiny min}}\!=\!2^{1/6}$, for simplicity, by defining $r_c\!\equiv\!x_c/x_{\mbox{\tiny min}}$. The influence of the interaction parameter $r_c$ on mechanical properties and mode-I (tensile) fracture has been investigated in~\cite{richard2021brittle}. Here, we fix $r_c\!=\!1.2$, which allows to efficiently prepare large samples that are brittle in the low temperature limit. The short range nature of the potential also enables one to efficiently study a wide range of strain-rates and finite size effects. A corollary of this choice of $r_c$ value is that Poisson's ratio of our computer glasses, $\nu\!<\!0.3$, is smaller than the experimental one, $\nu\!\simeq\!0.36$. A more thorough study of the influence of $\nu$ on the competition between shear-driven plasticity and dilation-driven void formation is currently underway, and hopefully will be reported on elsewhere.

The computer glass samples are cylindrical rods of length $L_0$ and diameter $D_0$. The initial configuration is created following the casting procedure put forward in~\cite{shi2010size}. A liquid is cast into an amorphous cylinder mold with a repulsive potential. The interaction between the mold and liquid corresponds to the same inverse-power-law (IPL) potential as developed in~\cite{lerner2019mechanical}. The system is then quenched at a rate of $\dot{Q}$ until the temperature reaches $T\!=\!0.05$, which is well below the estimated simulation glass temperature $T_{\rm g}\!\simeq\!0.35$ (cf.~Fig.~2c in the main text). Here, the quench rate is kept fixed at $\dot{Q}\!=\!10^{-4}$. The system is then brought to mechanical equilibrium via an energy minimization. The typical density at zero pressure in our samples, after the quench, is $\rho\!=\!N/V\simeq0.52$. Unless specified otherwise, we employ simulational units, where energies are expressed in terms of $\varepsilon$, temperature in terms of $\varepsilon/k_B$ and lengths in terms of the typical inter-particles distance $d_0\!=\!\rho^{-1/3}$. In order to highlight the differences between length and time scale accessible to atomistic simulations compared to laboratory experiments, we convert stress, length, strain rate into GPa, nm and s$^{-1}$ units, respectively. In practice, we employ the following conversions $\sigma\=2.352\,\AA$, $\varepsilon\=0.1$ eV, and $m\=91$ amu, following~\cite{puaduraru2010computer}.

Mechanical tests are performed in the NVT ensemble using a Nose-Hoover thermostat with time scale $\tau\!=\!1$. We verified that this choice allows to control the temperature up to the largest strain-rate employed in this study. The deformation is controlled by moving one end of the sample at a constant speed $v$, while keeping the other end fixed. The engineering strain-rate is defined as $\dot{\epsilon}\=v/L_0$.

In all of our simulations $L_0\!=\!35.3$ nm was used, and $D_0$ was varied from $11.8$ to $47$ nm. Accordingly, the number of particles in the system spans the range from $N\!\simeq\!334$ K (for $D_0\!=\!11.8$) to $N\!\simeq\!5.34$ M (for $D_0\!=\!47$). The system sizes employed in this study are comparable to state-of-the-art numerical studies on tensile failure in amorphous materials~\cite{bonfanti2018damage}. The smallest system allows to probe a wide range of strain-rates (about 3 orders of magnitude), while keeping the pulling speed at $v\!=\!5.10^{7}s^{-1}$ for $D_0\!>\!11.8$ and vary $\sigma_{\mbox{\tiny UTS}}$ by varying the temperature. All of the large-scale simulations are performed with the MPI domain decomposition implemented in the LAMMPS package~\cite{plimpton1995fast}.

\section{Detecting shear-driven plasticity and dilation-driven void formation at the particle level} 

\begin{figure}[h!]
\includegraphics[width = 0.9\textwidth]{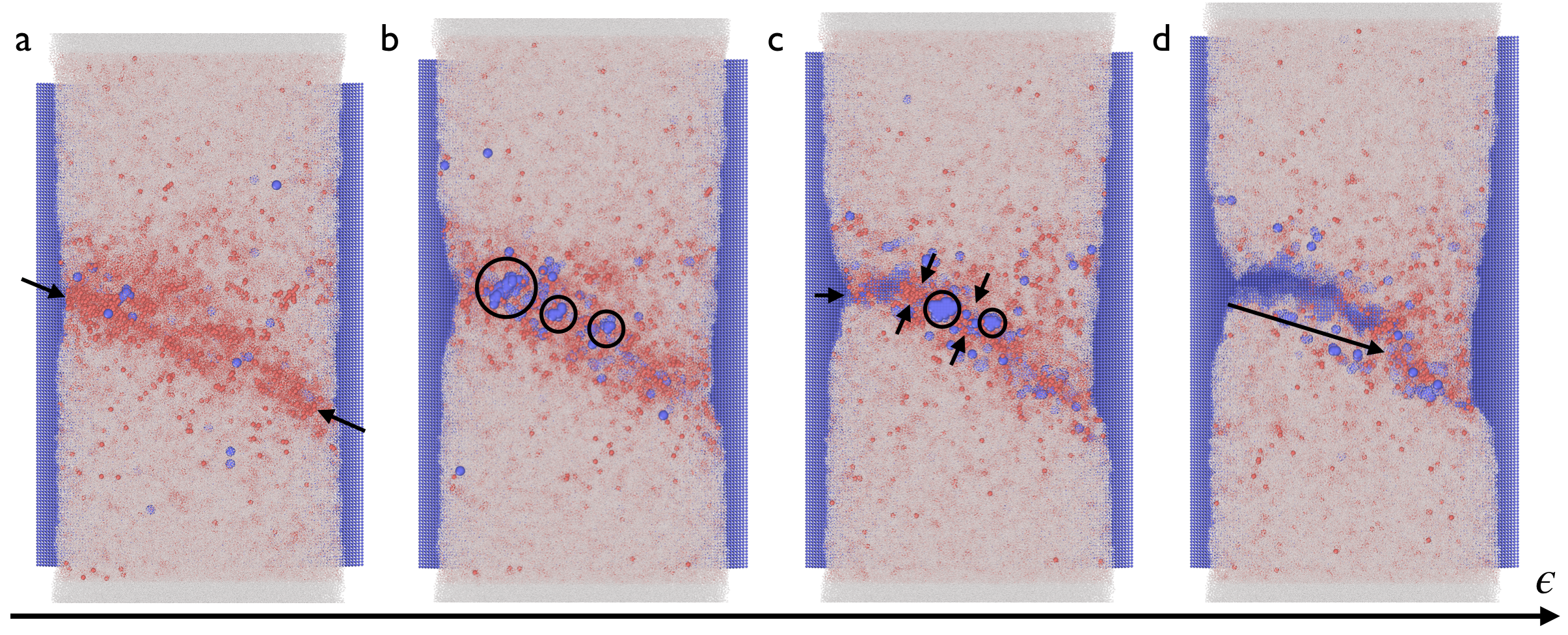}
\caption{\footnotesize An example for the evolution of shear-driven plasticity and dilation-driven void formation on the way to failure. (a) Strain localization (indicated by the arrows) at $\sim\!45^\circ$ to the tensile axis and the accompanying nucleation of small, sparsely distributed, voids (blue particles). (b) Voids grow into cavities inside the shear-band (highlighted by the circles). (c) The activation of shear-driven plasticity between cavities (pairs of arrows) and a cavity that percolates through the free surface (horizontal arrow). (d) A crack propagates via the coalescence of multiple cavities (the arrow indicates the crack propagation direction). Blue and red particles indicate shear-driven plasticity and voids, respectively. We distinguish particles belonging to the cluster that surrounds the sample from bulk voids using small and large radii, respectively.}
\label{fig:sm_void}
\end{figure}

In this study, shear-driven plasticity is monitored using the $D^2_{\rm min}$ field defined in~\cite{falk1998dynamics}. The $D^2_{\rm min}$ field provides a measure of the local (particle scale) non-affine deformation computed between 2 glass configurations separated by a strain interval $\Delta\epsilon$. In our analysis, we set $\Delta\epsilon\!=\!0.005\%$. Particles with a large non-affine deformation, corresponding to $D^2_{\rm min}\!>\!1$, are rendered in red. Here, we do not track cumulative shear-driven plasticity, i.e., a particle is marked only once, even if it took part in multiple plastic events. Dilation-driven void formation is monitored in the following manner: we place ``ghost/fictive particles'' on a regular cubic grid everywhere inside and around the rod, see blue particles in Fig.~\ref{fig:sm_void}, at a density identical to that of the computer glass. Ghost particles that do not overlap with real particles for at least a distance $d_0\!=\!\rho^{-1/3}$ are considered as voids. We are only interested in voids that are forming within the bulk (blue particles exposed near the rod's boundaries correspond to shear-driven plasticity, e.g., when a neck is formed, and hence are not voids). Consequently, we perform a cluster analysis to remove voids that percolate with ghost particles that surround the sample. Bulk voids are rendered as blue particles in Fig.~\ref{fig:sm_void}. Note that the blue regions can be larger than a single void, i.e., to correspond to a cluster of voids, which we term a cavity. Finally, we compute the components of the second order gyration tensor associated with the largest cavity according to  $S_{\alpha\beta}\!=\!\frac{1}{n_c}\sum_i^{n_c}r_\alpha^{(i)}r_\beta^{(i)}$, with $r_\alpha^{(i)}$ being the $\alpha$ component of the $i^{\hbox {\tiny th}}$ particle position and $n_c$ the number of voids in the cavity. We then extract the cavity (cluster) size $\lambda$ along the loading direction as $\lambda=\sqrt{S_{xx}}$, with $S_{xx}$ being the tensor component along the x-axis.

The above procedure and analysis are demonstrated in Fig.~\ref{fig:sm_void} for a sample in which failure is shear-banding mediated (i.e., for a low ultimate tensile strength), for different stages (strains) in the process. In Fig.~\ref{fig:sm_void}a, an oblique ($\sim\!45^\circ$ to the tensile axis) shear-band is observed (indicated by the arrows), accompanied by a small number of voids. With increasing strain, we observe (cf.~Fig.~\ref{fig:sm_void}b) the nucleation of voids inside the shear-band and their growth into larger cavities (indicated by the circles). Further increasing the strain, intense shear-driven plasticity is observed between cavities (indicated by the two pairs of arrows in Fig.~\ref{fig:sm_void}c), which promotes further growth. In parallel, voids percolate through the free surface (indicated by the horizontal arrow in Fig.~\ref{fig:sm_void}c). Finally, cavities coalesce to form a catastrophic crack that propagates through the system, as indicated by the long arrow in Fig.~\ref{fig:sm_void}d.  

\section{Ultimate tensile strength and equivalent states}

\begin{figure}[h!]
\includegraphics[width = 0.9\textwidth]{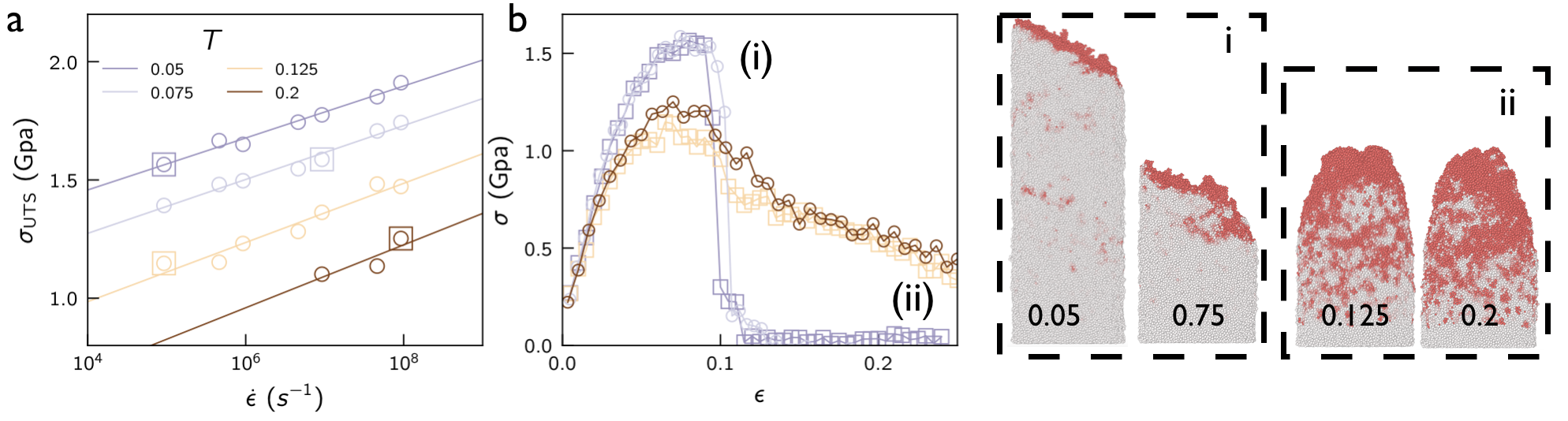}
\caption{\footnotesize (a) Ultimate tensile stress $\sigma_{\mbox{\tiny UTS}}$ vs.~ $\log(\dot{\epsilon})$ for different temperatures (as indicated in the legend). The solid lines are linear fits to $\sigma_{\mbox{\tiny UTS}}\!=\!a\log(\dot{\epsilon})+b$. (b) $\sigma(\dot{\epsilon})$ for two pairs of ``equivalent states'', i.e., strain-strain curves featuring the same $\sigma_{\mbox{\tiny UTS}}$ (indicated by empty squares in panel (a)), though obtained at different temperatures and strain-rates. (c) The corresponding postmortem samples are shown, where red particles indicate shear-driven plastic deformation.}
\label{fig:sm_equi}
\end{figure}

In Fig.~\ref{fig:sm_equi}a, we show the ultimate tensile strength $\sigma_{\mbox{\tiny UTS}}$ as a function of the logarithm of the strain-rate $\dot{\epsilon}$, for various temperatures (see legend). We find a linear relation between $\sigma_{\mbox{\tiny UTS}}$ and $\log(\dot{\epsilon})$, which is indicative of thermal activation processes. Next, we select two pairs of stress-strain curves $\sigma(\epsilon)$, one pair --- marked as (i) in Fig.~\ref{fig:sm_equi}b --- at a high $\sigma_{\mbox{\tiny UTS}}$ and another --- marked as (ii) in Fig.~\ref{fig:sm_equi}b --- at a low $\sigma_{\mbox{\tiny UTS}}$ obtained at different temperatures and strain-rates, yet sharing the same $\sigma_{\mbox{\tiny UTS}}$ (indicated by empty squares in Fig.~\ref{fig:sm_equi}a). In Fig.~\ref{fig:sm_equi}b, we superpose the four $\sigma(\epsilon)$ curves, which reveal nearly overlapping functions for pairs featuring the same ultimate tensile strength $\sigma_{\mbox{\tiny UTS}}$. These results demonstrate that $\sigma_{\mbox{\tiny UTS}}$ can indeed be used as to parameterize the crossover between necking- and shear-banding-mediated failure. Finally, Fig.~\ref{fig:sm_equi}c reveals the corresponding similarity in the postmortem failure patterns. 

\section{Failure pathways and void coalescence} 

\begin{figure}[h!]
\includegraphics[width = \textwidth]{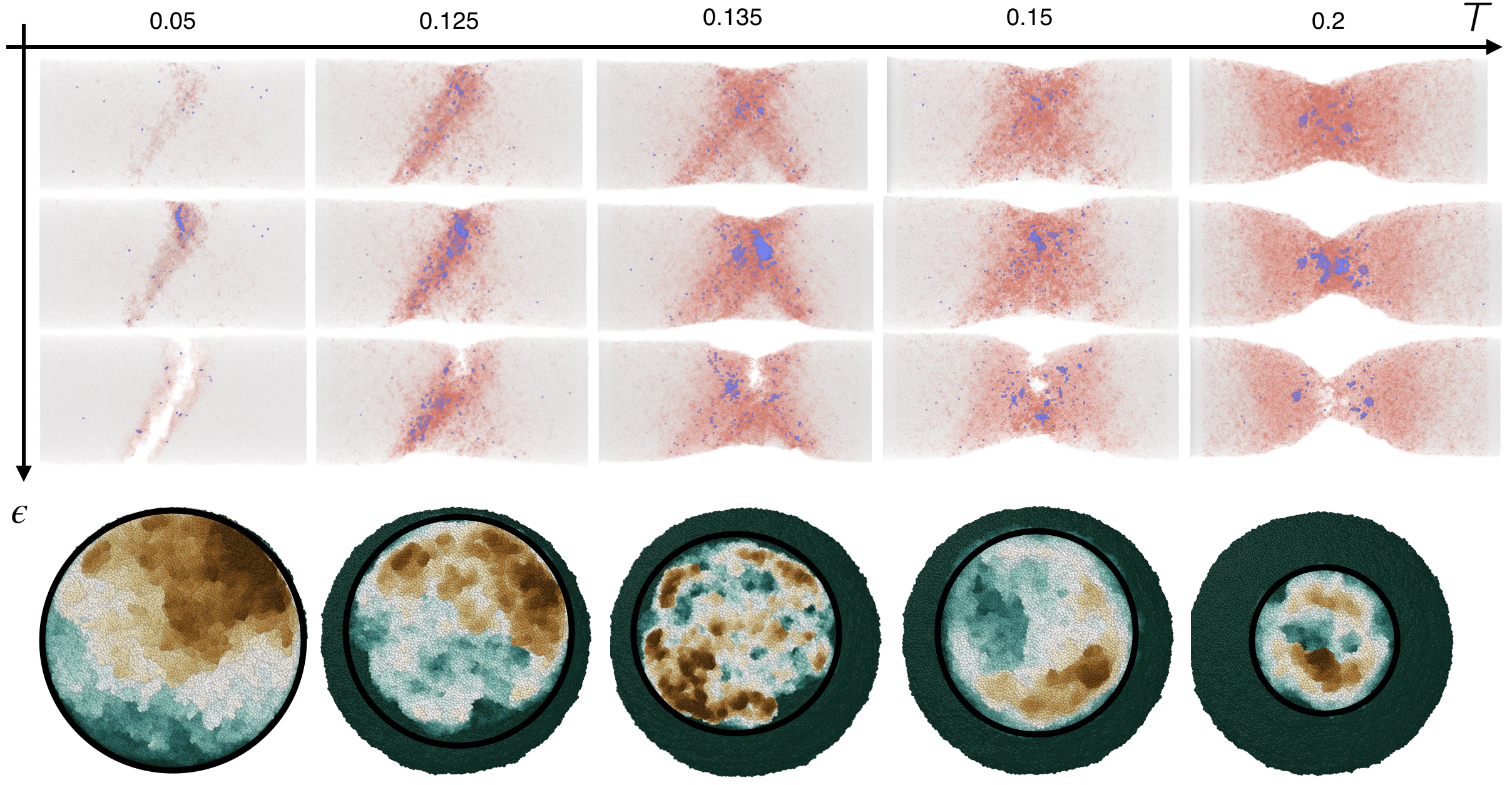}
\caption{\footnotesize Shear-driven plasticity (red) and dilation-driven void formation (blue) across a given failure pathway (increasing strain $\epsilon$, vertical arrow) for various temperatures (horizontal arrow). Bottom row shows the corresponding postmortem fractography, where colors represent the depth (from green to brown) and the black circles mark the fractured surface (the surrounding green region corresponds to the reduction in the cross-sectional area associated with neck formation).}
\label{fig:sm_pathway}
\end{figure}

In Fig.~\ref{fig:sm_pathway}, we present additional results showing the failure pathway (evolution with strain $\epsilon$, vertical arrow) for various temperatures (horizontal arrow), and the corresponding postmortem fractography. As discussed in the main text, the transition between shear-banding- and necking-mediated failure is associated with a transition from micro cavitation within a pre-nucleating shear-band in the former to void growth and coalescence of multiple cavities in the latter, leading to fracture at a much lower tensile stress. Consistently with experiments, albeit as significantly smaller length scales, fractography of the as-failed computer glasses clearly reveals the existence of voids with increasing temperature, see the bottom row of Fig.~\ref{fig:sm_pathway}. Voids can grow to larger sizes with increasing temperature (e.g., compared the middle panel to the rightmost one therein). 

\begin{figure}[h!]
\includegraphics[width = 0.9\textwidth]{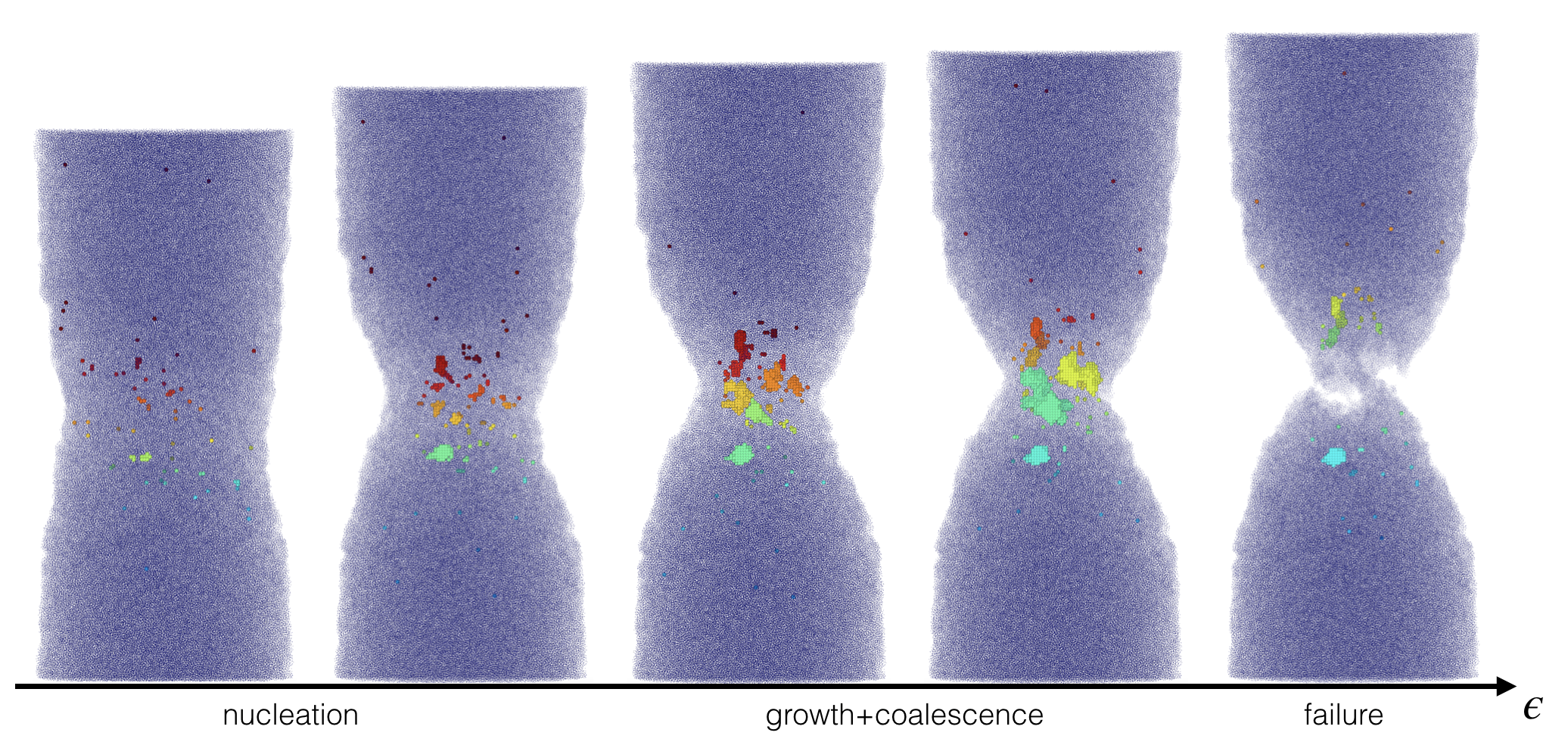}
\caption{\footnotesize Nucleation, void growth and coalescence, and failure as a function of $\epsilon$ at $T\!=\!0.2$, see text for details. Spatially independent clusters are rendered with different colors.}
\label{fig:sm_growth}
\end{figure}

Finally, in Fig.~\ref{fig:sm_growth}, we present an example of voids nucleation, growth and coalescence in the necking-mediated failure regime. Failure is first initiated by the nucleation of sparsely distributed voids (two leftmost snapshots in Fig.~\ref{fig:sm_growth}). With increasing strain $\epsilon$, these voids grow into cavities and start to coalesce, reaching a critical size $\lambda_{\rm F}$ (next two snapshots in Fig.~\ref{fig:sm_growth}), leading to catastrophic crack propagation and failure (rightmost snapshot Fig.~\ref{fig:sm_growth}). Residual cavities located inside the glasses, away from the fractured surface, are observed.

\section{Cluster distribution during shear-banding} 

\begin{figure}[h!]
\includegraphics[width = \textwidth]{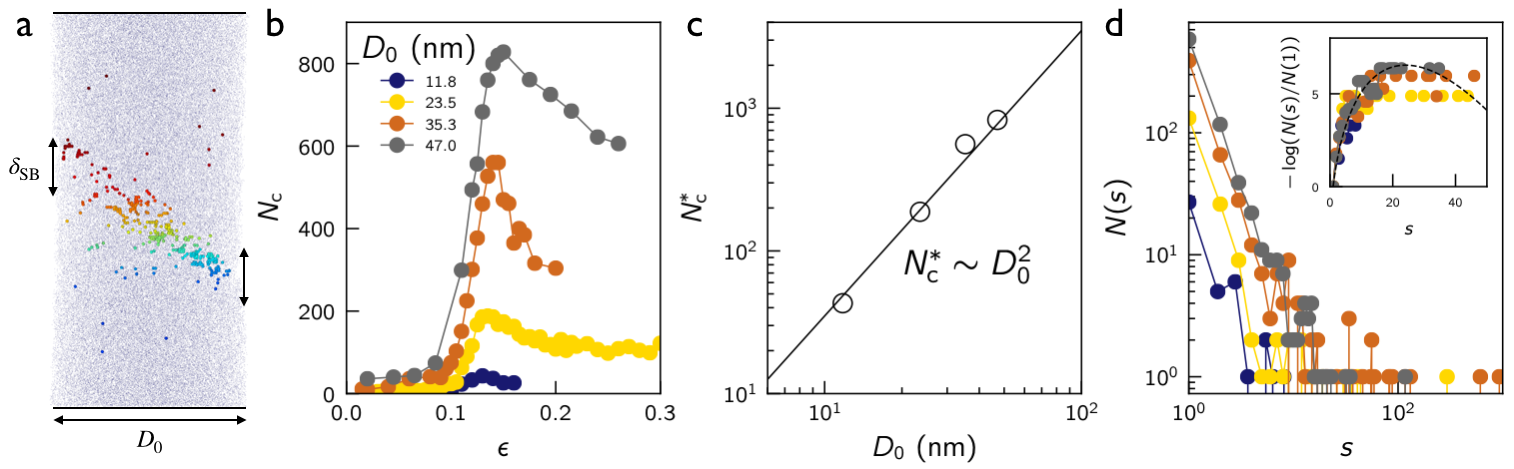}
\caption{\footnotesize (a) A snapshot of a sample during shear-banding and the accompanying cavitation at $T\!=\!0.1$. Spatially independent void clusters are rendered with different colors. The vertical double arrows indicate the shear band width $\delta_{\rm SB}$ and the horizontal double arrows indicates the pillar diameter $D_0$. (b) The number of independent clusters $N_{\rm c}$ is plotted as a function of $\epsilon$ for various $D_0$. (b) The largest number of clusters, $N_{\rm c}^*$, is plotted as a function of $D_0$. The solid line corresponds to the scaling $N_{\rm c}^*\!\sim\! D_0^2$. (d) The number distribution $N(s)$ of clusters of size $s$ just prior failure. The inset shows $-\log(N(s)/N(1))$ as a function of $s$. The dashed line, which corresponds to a free energy profile following $\Delta{F}(s)\!=\! -a s + bs^{2/3}$, is a guide to the eye.}
\label{fig:sm_cluster}
\end{figure}

During shear-banding-mediated failure, we observe a homogeneous nucleation of voids that primarily localize along the band, see snapshot in Fig.~\ref{fig:sm_cluster}a. This result clearly indicates that shear-driven plasticity induced softening lowers the nucleation barrier for cavitation. As shown in Fig.~\ref{fig:sm_cluster}b, the number of independent clusters $N_{\rm c}$ grows rapidly with strain, as failure is approached. The maximal number of clusters $N_{\rm c}^*$ (as a function of strain) grows with the pillar diameter $D_0$. We expect the number of clusters to be controlled by the volume of the shear band $V_{\rm SB}\!\sim\! \delta_{\rm SB}D_0^2$, with $\delta_{\rm SB}$ the typical band width. As we only observe a single shear band, which does not vary much in width across our diameter range, we expect $N_{\rm c}^*$ to scale as $\sim D_0^2$. This prediction is confirmed in Fig.~\ref{fig:sm_cluster}c. In Fig.~\ref{fig:sm_cluster}d, we plot the number $N(s)$ of cluster of size $s$. We find that the probability to observe large clusters is small. For homogeneous cavitation, we expect $N(s)\=N(1)\exp{\left(-\Delta{F}(s)/k_{\rm B}T\right)}$, with $N(1)$ the number of voids ($s\=1$) and $\Delta{F}(s)$ being the free energy cost to form a cluster of size $s$, at a given stress state. We plot $-\log(N(s)/N(1))$ in the inset of Fig.~\ref{fig:sm_cluster}d and find that distributions for different pillar diameters nicely overlap with each other, indicating a single nucleation barrier of the form $\Delta{F}(s)\=-a s + bs^{2/3}$, which reflects a competition between a volume and surface contribution, as in classical nucleation theory. Nucleation barriers are of the order of a few $k_{\rm B}T$, which explains the rapid growth that leads to abrupt failure. These observations are fully consistent with the picture put forward in~\cite{wright2003free}.

\end{document}